\documentclass[12pt]{article}

\usepackage{graphicx}
\graphicspath{{plots/}}
\usepackage{amsmath, natbib, url, hyperref, booktabs}
\usepackage{amssymb, amsfonts, color, bm, float, bbm, multirow}
\usepackage{amsthm} 
\usepackage{rotating}
\usepackage{microtype}
\usepackage{algorithm, algorithmic} 
\usepackage{enumitem} 
\setlist{nolistsep} 
\newtheorem{result}{Result}

\DeclareMathOperator{\E}{\mathbb{E}}

\DeclareMathOperator{\trace}{tr}

\DeclareMathOperator{\diag}{diag}
\floatstyle{ruled}

\begin{document}

\title{Bayesian experimental design without posterior calculations: an adversarial approach}
\author{Dennis Prangle\\
       School of Mathematics, University of Bristol, UK\\
       Sophie Harbisher, Colin S Gillespie \\
       School of Mathematics Statistics and Physics, Newcastle University, UK}
\date{}

\maketitle

\bigskip

\begin{abstract}
Most computational approaches to Bayesian experimental design
require making posterior calculations repeatedly for a large number of potential designs and/or simulated datasets.
This can be expensive and prohibit scaling up these methods to models with many parameters, or designs with many unknowns to select.
We introduce an efficient alternative approach without posterior calculations,
based on optimising the expected trace of the Fisher information, as discussed by Walker (2016).
We illustrate drawbacks of this approach, including lack of invariance to reparameterisation
and encouraging designs in which one parameter combination is inferred accurately but not any others.
We show these can be avoided by using an adversarial approach:
the experimenter must select their design while a critic attempts to select the least favourable parameterisation.
We present theoretical properties of this approach and show it can be used with
gradient based optimisation methods to find designs efficiently in practice.
\end{abstract}

\section{Introduction}

Selecting a good design for an experiment can be crucial to extracting useful information and controlling costs.
Applications include medical interventions \citep{Amzal:2006}, epidemic modelling \citep{Cook:2008}, pharmacokinetics \citep{Ryan:2014, Overstall:2017} and ecology \citep{Gillespie:2019}.
In modern applications it is increasingly feasible to take a large number of measurements --
e.g.~placing sensors \citep{Krause:2009} or making observations in a numerical integration problem \citep{Oates:2020} --
or make other complex designs -- e.g.~selecting a time series of chemical input levels to a synthetic biology experiment \citep{Bandiera:2018}.
Therefore finding high dimensional designs is an increasingly relevant task.

We focus on the Bayesian approach to optimal experimental design, which takes into account existing knowledge and uncertainty about the process being studied before the experiment is undertaken.
In this framework an \emph{experimenter} must select a design.
They then receive some utility based on the outcome of the experiment.
The aim is to select the design optimising expected utility given the experimenter's prior beliefs.

Most utility functions in Bayesian experimental design require posterior calculations, such as evaluating the evidence.
Optimisation of the design requires these calculations to be repeated for a large number of potential designs and simulated datasets.
This can be expensive and prohibit scaling up these methods to models with many parameters, or designs with many unknowns to select.
Some sophisticated approaches have been developed, including converting the optimisation problem into Markov chain Monte Carlo on the joint space of designs, parameters, and observations \citep{Muller:1999}, and using variational inference to learn approximate surrogate posteriors \citep{Foster:2019, Foster:2020}, but these still effectively solve inference problems and hence remain costly.
(We comment on other related approaches in Section \ref{sec:related}.)

This paper presents approaches which avoid the need for posterior inference.
First we consider a utility function based on the trace of the Fisher information, which is often available in a simple closed form.
\cite{Walker:2016} presents an information theoretic justification for this utility.
One contribution of this paper is to show that it also emerges naturally from the decision theoretic framework of \cite{Bernardo:1979}.
Our derivation is based on judging the quality of a parameter estimate through the Hyv\"arinen score \citep{Hyvarinen:2005}, rather than through the logarithmic score as in \cite{Bernardo:1979}.
We demonstrate that it is straightforward to optimise the resulting expected utility using recent developments in stochastic gradient optimisation \citep{Kingma:2015} and automatic differentiation \citep{Baydin:2017}.
Compared to existing methods, this approach is fast and scales easily to higher dimensional designs.
However, a drawback is that the method can converge to poor local maxima.
We show how a second stage of optimisation similar to that of \cite{Overstall:2017} can often be used to find the overall optimal design.

A more fundamental limitation of the above approach is that it sometimes produces poor designs in practice e.g.~requiring all observations to occur at a single time point.
We provide an explanation:
optimising this expected utility encourages designs giving accurate inference of one linear combination of the parameters but not necessarily others.
Furthermore this utility is not invariant to reparameterisation:
this can alter which parameter combinations are most rewarding to infer accurately.

We address both these issues by introducing an adversarial approach.
We propose a game theoretic framework in which, as before, the experimenter chooses a design to optimise their expected utility.
Now there is also a \emph{critic} who selects a linear transformation of the parameters.
We investigate the optimal designs in this framework under the game theoretic solution concept of subgame perfect equilibrium, and prove they are invariant to reparameterisation.
The presence of the critic also encourages designs not to neglect the posterior accuracy of any parameter combination:
if one did then the critic could choose a parameterisation concentrating on this weakness.
We show it is possible to find optimal designs in this game theoretic framework using generic \emph{gradient descent ascent} methods, which have been much studied recently in the machine learning literature \citep[e.g.][]{Heusel:2017, Jin:2019}.

Below, we summarise our contributions in Section \ref{sec:contrib}
and related work in Section \ref{sec:related}.
In the remainder of the paper,
Section \ref{sec:background} presents background on Bayesian experimental design.
Section \ref{sec:DTtheory} presents results on the decision theoretic framework of \cite{Bernardo:1979}, and Section \ref{sec:GTtheory}, on our proposed game theoretic extension.
Proofs and other technical material are presented in the appendices.
Section \ref{sec:opt} discusses details of gradient based optimisation for both approaches.

We illustrate our methods on a simple Poisson model where optimal designs can be derived analytically (Section \ref{sec:illustration}).
We provide a detailed simulation study on a pharmacokinetic model (Section \ref{sec:pk}),
showing our adversarial approach produces a sensible design and is at least 10 times faster than competing methods.
We also present a geostatistical regression example, where hundreds of design choices can be optimised in under a minute (Section \ref{sec:geostat}).
Code for these examples is available at
\url{https://github.com/dennisprangle/AdversarialDesignCode}.
All examples were run on a desktop PC with 12 CPU cores.
Our conclusion, Section \ref{sec:discussion}, summarises our findings and recommendations to implement our methods.
It also discusses limitations of our work, and future research directions to address these, including a discussion of the intractable Fisher information case (detailed further in the appendices).

\subsection{Contributions} \label{sec:contrib}

Our main contribution is a faster approach to Bayesian experimental design using the Fisher information.
Where the Fisher information is available in closed form,
our approach outperforms existing methods (Section \ref{sec:pk})
and scales easily to designs with hundreds of design choices (Section \ref{sec:geostat}).

In addition, we contribute by extending the decision theoretic framework for Bayesian experimental design of \cite{Bernardo:1979} to allow other proper scoring rules in addition to logarithmic score (Section \ref{sec:DTframework}).
We also show that using the Hyv\"arinen score in the decision theoretic framework results in a utility based on the trace of the Fisher information,
referred to as Fisher information gain or FIG in the paper (Section \ref{sec:DTres}).
This provides a decision-theoretic justification to a measure that was previously suggested by \cite{Walker:2016} based on information theoretic arguments.
We explore the limitations of FIG (Section \ref{sec:FIGprop})
and address them by extending the decision theoretic approach to a game theoretic approach,
and show the relevant solution concept gives more intuitively informative designs (Section \ref{sec:GTtheory}).
We also show that the game theoretic solution can easily be obtained in practice using gradient-based minimax optimisation.

Finally, we provide new insight into two long-standing questions in Bayesian experimental design.
Firstly: what counts as a Bayesian utility function?
(See Sections \ref{sec:FIGprop} and \ref{sec:properties} under the heading ``Bayesian justification''.)
Secondly: which of several possible Bayesian generalisations of $D$-optimality should be preferred?
(See Section \ref{sec:properties} under the heading ``Link to $D$-optimality''.)

\subsection{Related work} \label{sec:related}

As discussed above, our theoretical contribution builds on
\cite{Bernardo:1979}'s decision theoretic framework for Bayesian experimental design,
and also on \cite{Walker:2016}'s justification for the use of a utility based on the trace of the Fisher information.
We also discuss \cite{Overstall:2020}'s comments on this utility later.

Gradient-based optimisation methods for experimental design have been explored previously.
\cite{Pronzato:1985} optimise the expected determinant of the Fisher information using analytically derived gradients.
\cite{Huan:2013, Huan:2014} optimise expected Shannon information using gradients
(either derived analytically or based on finite differences)
for a biased numerical approximation to the utility.
\cite{Foster:2019} use a first stage of variational inference to learn an approximate posterior, and then use this to produce a surrogate expected utility function which is optimised by gradient-based methods in a second stage.

\cite{Foster:2020} and \cite{Kleinegesse:2020} propose jointly optimising a design and some tuning parameters defining a lower bound on the expected Shannon information gain utility (this utility is discussed in Section \ref{sec:utilities}).
This can be framed as a single optimisation problem, allowing for easy implementation using stochastic gradient or Bayesian optimisation methods.
Producing a tight lower bound is equivalent to, or closely related to, being able to perform exact Bayesian inference for the optimal design.

\cite{Overstall:2017} propose an alternative coordinate ascent approach which loops over the components of the design, updating each in turn.
To perform an update, designs 
are selected from the one dimensional search space (in which only the current component is updated)
and Monte Carlo estimates of expected utility calculated. 
A Gaussian process is fitted to the expected utility estimates and used to propose an improved value for the design component under consideration.
This is accepted or rejected based on a Bayesian test of whether it improves expected utility, using a large number of simulations under the current and proposed designs.



Many other algorithms have been proposed for Bayesian experimental design.
An influential method of \cite{Muller:1999} performs optimal design using Markov chain Monte Carlo.
\citealp{Amzal:2006} and \citealp{Kuck:2006} extend this approach to use sequential Monte Carlo.
In other approaches, \cite{Ryan:2014} look at high dimensional designs with a low dimensional parametric form,
\cite{Price:2018} use evolutionary algorithms, and
\cite{Gillespie:2019} search a discrete grid of designs.

\section{Background} \label{sec:background}

Optimal experimental design concerns the following problem.
An experimenter must select a design $\tau$.
The experiment produces data $y$ with likelihood $f(y|\theta; \tau)$,
where $\theta$ is a vector of parameters.
The goal is to select the design which optimises some notion of the quality of the experiment,
typically based on its informativeness and its cost.
We mostly, but not exclusively, concentrate on the case where
$\tau \in \mathcal{T} \subseteq \mathbb{R}^d$
and $\theta \in \Theta \subseteq \mathbb{R}^p$
for some closed sets (under the Euclidean metric) $\mathcal{T}$ and $\Theta$,
and where $y$ is a vector of $n$ observations.
Designs $\tau$ of this form often represent times or locations for measurements to be taken.
In this case $\tau$ can be seen as a set of \emph{design points}, $\tau_1, \tau_2, \ldots, \tau_d$.

This section reviews relevant details of the statistical background.
First Sections \ref{sec:Bayes} and \ref{sec:FIM} give some necessary definitions on Bayesian statistics and Fisher information.
Section \ref{sec:BED} describes Bayesian experimental design, which is based on maximising expected utility.
Finally, Section \ref{sec:utilities} summarises some common utility functions.

\subsection{Bayesian framework} \label{sec:Bayes}

We work in the Bayesian framework and introduce a prior density $\pi(\theta)$ for $\theta$.
We will often make use of the posterior density and the prior predictive density (or \emph{evidence}) for $y$.
In our setting both depend on the experimental design $\tau$,
\begin{align}
\pi(\theta | y; \tau) &= \pi(\theta) f(y | \theta; \tau) / \pi(y; \tau), \label{eq:posterior} \\
\pi(y; \tau) &= \E_{\theta \sim \pi(\theta)}[f(y | \theta; \tau)] \label{eq:prior predictive}.
\end{align}
The prior and model define a joint density\footnote{
For discrete $y$ a density with respect to a product of Lebesgue and counting measures can be used.},
\begin{equation} \label{eq:joint}
\pi(\theta, y; \tau) = \pi(\theta) f(y | \theta; \tau)
= \pi(\theta | y; \tau) \pi(y; \tau).
\end{equation}

\subsection{Fisher information} \label{sec:FIM}

We will make frequent use of the Fisher information matrix (FIM) for $\theta$,
\begin{equation} \label{eq:FisherInfo}
\mathcal{I}_\theta(\theta; \tau) = \E_{y \sim f(y | \theta; \tau)} [ u(y, \theta; \tau) u(y, \theta; \tau)^T ],
\end{equation}
which is based on the score function,
\begin{equation} \label{eq:score}
u(y,\theta;\tau) = \nabla_\theta \log f(y|\theta; \tau).
\end{equation}
The subscript in $\mathcal{I}_\theta$ denotes which variable is used for differentiation in \eqref{eq:score}.
Often this is obvious from the context so the subscript will be dropped.
We will also sometimes omit the dependence on $\tau$ where this is not relevant.
We will focus on models where $u$ and $\mathcal{I}$ are well defined.

\paragraph{Examples}

The FIM for exponential family models is based on the variance of the sufficient statistics (see e.g.~\citealp{Lehmann:2006}) which is often available in closed form.
Two examples we use in this paper are:
\begin{itemize}
\item
The Poisson distribution, $y \sim Poisson(\phi)$.
Here $\mathcal{I}(\phi) = 1/\phi$.
\item
The multivariate normal distribution with known variance,
$y \sim N(\mu, \Sigma)$.
Here $\mathcal{I}(\mu) = \Sigma^{-1}$.
\end{itemize}


\paragraph{Reparameterisation}

Consider a probability model with parameters $\theta$,
and a function $\phi(\theta)$
producing an alternative vector of parameters $\phi$
(which may be shorter or longer than $\theta$).
Let $J(\phi)$ be the Jacobian
i.e.~the matrix whose row $i$ column $j$ entry
is $\frac{\partial \phi_i}{\partial \theta_j}$.
Then it follows from \eqref{eq:FisherInfo} that
\begin{equation} \label{eq:reparam}
\mathcal{I}_\theta(\theta) = J(\phi)^T \mathcal{I}_\phi(\phi) J(\phi).
\end{equation}
An application of this result is to 
the model $y \sim N(x(\theta,\tau), \Sigma)$.
Using \eqref{eq:reparam},
\begin{equation} \label{eq:mvn_info}
\mathcal{I}_\theta(\theta) = J(x)^T \mathcal{I}_x(x) J(x)
= J(x)^T \Sigma^{-1} J(x).
\end{equation}

\subsection{Bayesian experimental design} \label{sec:BED}

The Bayesian approach to experimental design involves selecting a function $\mathcal{U} = \mathcal{U}(\tau, \theta, y)$,
giving the utility of choosing design $\tau$ given observations $y$ and true parameters $\theta$.
(As we shall see, many choices of $\mathcal{U}$ do not depend on all these possible inputs.)
We try to maximise the \emph{expected utility} of $\tau$ i.e. the prior predictive utility
\begin{equation} \label{eq:expected utility}
\mathcal{J}(\tau) = \E_{(\theta,y) \sim \pi(\theta, y; \tau)}[\mathcal{U}(\tau, \theta, y)].
\end{equation}
See \cite{Chaloner:1995}, \cite{Atkinson:2007} and \cite{Ryan:2016} for comprehensive surveys of Bayesian experimental design.

\subsection{Utility functions} \label{sec:utilities}

Ideally a utility function could be specified for each application,
perhaps by eliciting preferences over different $(\tau,\theta,y)$ combinations from the experimenter \citep[e.g.][]{Wolfson:1996}.
However this is rarely feasible in practice.
Instead several generic choices of utility have been proposed.

\paragraph{Shannon information gain (SIG)}

This is a popular and well-motivated utility choice,
\begin{equation} \label{eq:SIG}
\mathcal{U}_{\text{SIG}}(\tau, \theta, y) = \log \pi(\theta | y; \tau) - \log \pi(\theta),
\end{equation}
introduced by \cite{Lindley:1956},
which is particularly relevant later in this paper.
Designs maximising the expectation of $\mathcal{U}_{\text{SIG}}$
have an appealing information theoretic interpretation:
they maximise expected reduction in Shannon entropy from prior to posterior.
Furthermore, \cite{Bernardo:1979} gave a decision theoretic derivation of $\mathcal{U}_{\text{SIG}}$ which we recap in Section \ref{sec:DTtheory}.
A helpful property of $\mathcal{U}_{\text{SIG}}$ is that it is \emph{reparameterisation invariant} i.e.~unchanged under a bijective transformation of $\theta$.
Thus the resulting designs are not affected by the choice of parameterisation.

In practice, optimising the expectation of \eqref{eq:SIG} is complicated by the need to evaluate the posterior density.
However, using \eqref{eq:posterior} the utility can be rewritten as
\begin{equation}
\mathcal{U}_{\text{SIG}}(\tau, \theta, y) = \log f(y | \theta; \tau) - \log \pi(y; \tau). \label{eq:SIG2}
\end{equation}
Optimisation now requires evaluation of a posterior summary: the \emph{log evidence} $\log \pi(y; \tau)$.
A common SIG estimate replaces $\pi(y; \tau)$ in \eqref{eq:SIG2} with a Monte Carlo approximation
\begin{equation} \label{eq:SIGMC}
\hat{\pi}(y; \tau) = \frac{1}{L} \sum_{\ell=1}^L f(y | \theta^{(\ell)}; \tau),
\end{equation}
where $\theta^{(\ell)}$ are independent prior samples.
A typical choice of $L$ is 1000 \citep{Overstall:2017},
which makes each utility evaluation somewhat computationally expensive.
Furthermore, a biased estimate of $\mathcal{U}_{\text{SIG}}$ is produced.
More efficient approaches are possible,
as outlined in Section \ref{sec:related}.
However the need for log evidence estimation remains a source of computational expense and approximation error.

\paragraph{Other utilities and Bayesian justification}

Many alternative utility functions to $\mathcal{U}_{\text{SIG}}$ are also used in practice e.g.~posterior precision and mean squared error between the posterior mean and true parameters.
See \cite{Chaloner:1995} and \cite{Ryan:2016} for detailed reviews.
Like SIG, both utilities just mentioned are also based on posterior calculations.
Indeed \cite{Ryan:2016} argue that for a utility to be ``fully Bayesian'', it must be a functional of the posterior,
and other utilities, such as scalar summaries of the FIM, are ``pseudo-Bayesian''.
(Throughout we use this definition of ``pseudo-Bayesian'',
but note some authors use the term differently e.g.~\citealp{Overstall:2020}.)
One contribution of this paper is to instead use a decision theoretic justification for which utilities to use in a Bayesian framework, which, surprisingly, provides support to some apparently pseudo-Bayesian utilities
(see discussion in Section \ref{sec:FIGprop}),
and then to further develop this into a game theoretic approach.

\section{Theory: decision theory approach} \label{sec:DTtheory}

This section explores a decision theoretic approach to underlie Bayesian experimental design.
Section \ref{sec:DTframework} presents the framework, which is taken from \cite{Bernardo:1979}.
Section \ref{sec:scoring rules} describes some background on \emph{scoring rules}.
Section \ref{sec:DTres} presents a novel theoretical result showing how the framework supports maximisation of expected utility for various classes of utility function derived from scoring rules.
This includes Shannon information gain, as shown by \cite{Bernardo:1979},
but also the trace of the FIM, as proposed by \cite{Walker:2016}.
We conclude in Section \ref{sec:FIGprop} by discussing advantages and disadvantages of the latter utility, and how these reveal limitations of the decision theory framework, motivating our modification in Section \ref{sec:GTtheory}.

\subsection{Decision theoretic framework} \label{sec:DTframework}

\cite{Bernardo:1979} proposed the following decision theoretic framework for Bayesian experimental design.
The experimenter selects a design, and then nature generates parameters $\theta$ from the prior $\pi(\theta)$ and observations $y$ from the likelihood $f(y|\theta;\tau)$.
The experimenter only observes $y$ and must now choose $a$, a density for $\theta$, receiving reward $\mathcal{R}(a,\theta)$.
In this section and Section \ref{sec:GTtheory} we will assume that:
\begin{enumerate}
\item[A1] The reward is the negative of a \emph{strictly proper scoring rule}, as defined in Section \ref{sec:scoring rules}.
\end{enumerate}
Throughout this section $\mathcal{S}(a,\theta) = -\mathcal{R}(a,\theta)$ denotes the scoring rule.

\subsection{Scoring rules} \label{sec:scoring rules}

A \emph{scoring rule} $\mathcal{S}(q, \theta)$ measures the quality of a distribution -- in this paper represented by its density $q(\theta)$ -- to model an uncertain quantity, given a realised value $\theta$.
Low scores represent a good match.
A scoring rule is \emph{strictly proper} if, given any $p(\theta)$,
$\E_{\theta \sim p(\theta)}[\mathcal{S}(q, \theta)]$ is uniquely minimised by $q=p$.
For more background on scoring rules see for example \cite {Gneiting:2007} and \cite{Parry:2012}.

Given a scoring rule, two related quantities are
\begin{align*}
\mathcal{H}[p(\theta)] &= \E_{\theta \sim p(\theta)}[\mathcal{S}(p(\theta), \theta)], &&
\text{(entropy of $p$)} \\
\mathcal{D}[p(\theta), q(\theta)] &= \E_{\theta \sim p(\theta)}[\mathcal{S}(q(\theta), \theta) - \mathcal{S}(p(\theta), \theta)]. \!&&
\text{(divergence from $p$ to $q$)}
\end{align*}
Appendix \ref{app:scoring} gives details of the entropy and divergence for two strictly proper scoring rules which will be used below: logarithmic score and Hyv\"arinen score \citep{Hyvarinen:2005}.
Hyv\"arinen score uses only the derivatives of $\log q(\theta)$,
so it can be calculated from unnormalised densities.

\subsection{Results} \label{sec:DTres}

Result \ref{res:DTmain} is a general result characterising solutions of the decision theoretic framework.
It uses the following extra assumption:
\begin{enumerate}
\item[A2]
Both
$\E_{\theta \sim \pi(\theta)}[ \, |S(\pi(\theta), \theta)| \, ]$
and, for any $\tau$,
$\E_{(\theta,y) \sim \pi(\theta, y; \tau)}[ \, |S(\pi(\theta|y;\tau), \theta)| \, ]$
are finite.
\end{enumerate}
\begin{result} \label{res:DTmain}
Assume A1 and A2.
Then the following are equivalent, in the sense of sharing the same set of optimal designs:
\begin{enumerate}
\item
The experimenter acts to maximise their expected reward.
\item
The experimenter selects $\tau$ to maximise the expectation, with respect to $\pi(\theta, y; \tau)$, of any of the following utilities:
\begin{align*}
\mathcal{U}_{\text{entropy}} &= -\mathcal{H}[\pi(\theta | y; \tau)], \\
\mathcal{U}_{\text{entropy diff}} &= \mathcal{H}[\pi(\theta)] - \mathcal{H}[\pi(\theta | y; \tau)], \\
\mathcal{U}_{\text{divergence}} &= \mathcal{D}[\pi(\theta | y; \tau), \pi(\theta)].
\end{align*}
\end{enumerate}
\end{result}
(Note that arguments of utilities are omitted in this section to simplify notation.)

The next result provides further equivalent utility choices for particular scoring rules.
\begin{result} \label{res:SIG/FIG}
For the logarithmic scoring rule, and assuming A2, maximising the expectation with respect to $\pi(\theta, y; \tau)$
of either $\mathcal{U}_{\text{divergence}}$ or $\mathcal{U}_{\text{SIG}}$, defined in \eqref{eq:SIG},
gives the same set of optimal designs.

\noindent For the Hyv\"arinen scoring rule the same is true for $\mathcal{U}_{\text{divergence}}$ and both of
\[
\mathcal{U}_{\text{FIG}} = ||\nabla_\tau \log f(y|\theta;\tau)||^2, \quad
\mathcal{U}_{\text{trace}} = \trace \mathcal{I}(\theta; \tau),
\]
assuming regularity condition A5\footnote{A5 implies A2, and also contains some other conditions.}, defined in Appendix \ref{app:scoring}.
\end{result}
Here $||\theta||$ represents the $L_2$ norm i.e.~$||x|| = \sqrt{x^T x}$.

The first part of Result \ref{res:SIG/FIG} provides a decision theoretic derivation of the Shannon information gain utility.
The argument used is essentially the same as that of \cite{Bernardo:1979}.
In an analogy to this, we refer to $\mathcal{U}_{\text{FIG}}$ in the second part of the result as \emph{Fisher information gain}\footnote{
Some earlier preprints of this paper used this name for $\mathcal{U}_{\text{trace}}$ instead.
Our usage here is a closer analogy to Shannon information gain, and matches that of \cite{Overstall:2020}.}.
We note that \cite{Walker:2016} proved directly that $\mathcal{U}_{\text{trace}}$ and the corresponding $\mathcal{U}_{\text{entropy diff}}$ and $\mathcal{U}_{\text{divergence}}$ all have the same expectations up to an additive constant,
and were therefore equivalent when used in experimental design.
The second part of Result \ref{res:SIG/FIG} shows that the same conclusion arises from a decision theoretic approach based on the Hyv\"arinen score.

\subsection{Fisher information gain properties} \label{sec:FIGprop}

Result \ref{res:SIG/FIG} supports maximising the expectation of $\mathcal{U}_{\text{FIG}}$, or equivalently of $\mathcal{U}_\text{trace}$,
\begin{equation} \label{eq:Jfig}
\mathcal{J}_{\text{FIG}}(\tau) = \E_{\theta \sim \pi(\theta)}[ \trace \mathcal{I}(\theta; \tau) ] = \trace \bar{\mathcal{I}}(\tau).
\end{equation}
where $\bar{\mathcal{I}}(\tau) = \E_{\theta \sim \pi(\theta)}[ \mathcal{I}(\theta; \tau) ]$.
We refer to this as the FIG approach to experimental design,
and here we discuss its properties.
We will see that despite computational advantages it has several undesirable properties, illustrated later in our examples.
Section \ref{sec:GTtheory} addresses these issues by generalising the decision theoretic framework.

\paragraph{Bayesian justification}

Since the definition of $\mathcal{U}_{\text{trace}}$ does not involve $y$,
it is not a functional of the posterior,
and therefore is pseudo-Bayesian under the terminology of \cite{Ryan:2016}
(discussed at the end of Section \ref{sec:utilities}.)
However, as pointed out by \cite{Walker:2016}, $\mathcal{J}_{\text{FIG}}$ also results from using utilities which \emph{are} functionals of the posterior -- e.g.~$\mathcal{U}_{\text{divergence}}$ in our notation --
so it can be regarded as fully Bayesian.
This shows the divide between pseudo-Bayesian and fully Bayesian approaches can be hard to clearly define, and motivates using a decision (/game) theoretic approach to do so.

\paragraph{Computational advantages}

The FIM is often available in a closed form which can easily be evaluated, and allows easy evaluation of gradients.
Then optimisation of $\mathcal{J}_{\text{FIG}}(\tau)$ is straightforward using standard stochastic optimisation methods, as described in Section \ref{sec:opt}.
In particular, the objective does not involve calculating the log evidence, which creates optimisation difficulties for $\mathcal{J}_{\text{SIG}}$, or any other use of explicit posterior inference.
The case where evaluation of the FIM is more complicated is discussed in Appendix \ref{sec:estimationMore}.

\paragraph{Lack of reparameterisation invariance}

The optimum of $\mathcal{J}_{\text{FIG}}(\tau)$ can change under a reparameterisation of $\theta$:
see Section \ref{sec:illustration} for a simple illustration.
This property is undesirable as the optimal design is affected by the seemingly irrelevant choice of what parameterisation is used.
In contrast, the design optimising expected Shannon information gain is reparameterisation invariant.

\paragraph{Intuitively uninformative designs}

By construction, optimal FIG designs maximise a particular definition of informativeness.
However these designs often have properties which seem ``uninformative'' in an intuitive sense.

One issue is that FIG designs often result in posterior distributions which are very diffuse for some linear combinations of parameters,
although they may be concentrated for others.
This is illustrated in Figure \ref{fig:pk_posterior}, where the design results in a long narrow posterior in which the marginals for two parameters are very similar to the prior marginals.
In contrast, a competing design produces posterior marginals which are concentrated around the true values for both parameters.

Also, FIG designs can have an excessive amount of \emph{replication}: repeated observations at the same time or location.
For example, Appendix \ref{app:pkSGD} shows a FIG design with \emph{all} observations at the same time.
Such repeated observations provide increasingly accurate inference on one parameter (or parameter combination) at the expense of the others.

An informal explanation for this behaviour is as follows.
The optimal FIG design maximises the trace of $\bar{\mathcal{I}}(\tau)$,
which equals the sum of its eigenvalues.
Often this sum is maximised when one eigenvalue is large and the others are small:
we provide an example in Section \ref{sec:illustration},
and also \citet{Overstall:2020} describes a linear model where this occurs.
In a Bayesian setting, $\mathcal{I}(\theta; \tau)$ is an approximation to posterior precision \cite[see e.g.][]{Vaart:2000}, so this corresponds to the typical posterior having one parameter combination which is accurately inferred at the expense of the others.
\cite{Overstall:2020} also points out that the optimal $\tau$ can produce a singular $\bar{\mathcal{I}}(\tau)$ matrix,
causing some statistical methods to break down,
and a reason for this is that $\mathcal{J}_{\text{FIG}}$ only uses terms on the diagonal of $\bar{\mathcal{I}}(\tau)$, allowing off-diagonal terms to make it singular.

\section{Theory: game theory approach} \label{sec:GTtheory}

The previous section provided a theoretical framework supporting the use of $\mathcal{U}_\text{trace}$,
but found it produced designs with several undesirable features,
including dependence on the choice of parameterisation and a diffuse marginal posterior for some linear combinations of parameters.
Here we modify the framework to produce a design which is robust to linear reparameterisation, which, as we shall see, encourages marginal posterior concentration for all parameter combinations.

Below, Section \ref{sec:GTframework} outlines our framework.
Section \ref{sec:GTdefs} reviews some game theory definitions,
which are used in Section \ref{sec:GTresults}
to characterise optimal designs supported by the framework,
including an adversarial variation on FIG which we refer to as ADV.
Section \ref{sec:properties} describes the properties of ADV and its advantages over FIG.

\subsection{Game theoretic framework} \label{sec:GTframework}

We propose the following game theoretic framework.
Initially, the experimenter selects a design $\tau$.
We introduce a \emph{critic} who now selects the parameterisation
by choosing an invertible matrix $A$ defining parameters $\phi = A^{-1} \theta$.
The experimenter must then select $a$, a density for $\phi$,
receiving a reward $\mathcal{R}(a,\phi)$.
The critic's reward is $-\mathcal{R}(a,\phi)$:
they aim to find the parameterisation for which the design does worst.
This completes the specification of a game,
as discussed further in Section \ref{sec:GTdefs}.

We continue to use assumption A1,
which now gives a scoring rule of $\mathcal{S}(a,\phi) = -\mathcal{R}(a,\phi)$.
We also introduce another assumption, which is discussed in Appendix \ref{app:det}:
\begin{enumerate}
\item[A3]
The critic is restricted to selecting $A$ with $\det A = 1$.
\end{enumerate}

Table \ref{tab:dec} in Appendix \ref{app:frameworks} summarises and contrasts both the decision-theoretic and game-theoretic frameworks.

\subsection{Game theory definitions} \label{sec:GTdefs}

We refer to $\tau,A,\theta,y,a$ as \emph{actions} of our game.
The game specifies a mapping from $\tau,A,\theta,y,a$ to real-valued rewards for the \emph{players}: the experimenter and critic.
The actions $\theta, y$ are random samples from specified distributions.
The remaining actions are chosen by the players according to \emph{decision rules},
$\tau, A(\tau), a(\tau, A, y)$.

In the remainder of this section we explore the set of \emph{subgame perfect equilibria} (SPEs): players act to optimise their expected reward under the assumption that later decisions will also do so.
In our setting, SPE is a condition on decision rules.
The first condition is that $a(\tau,A,y)$ must output some $a$ maximising the experimenter's expected reward given $(\tau,A,y)$.
The second condition is that $A(\tau)$ must output $A$ maximising the critic's expected reward given $\tau$ and $a(\tau,A,y)$.
Finally $\tau$ must maximise the experimenter's expected reward given $A(\tau)$ and $a(\tau,A,y)$.
Note that existence and uniqueness of decision rules meeting these conditions is not automatic.


SPEs are often argued \citep[see e.g.][]{Jin:2019}
to be an appropriate solution concept for games in which players act sequentially.
The alternative solution concept of \emph{Nash equilibria} is more appropriate for games with simultaneous actions.
For more background on game theory and solution concepts see e.g.~\cite{Osborne:1994}.

\subsection{Results} \label{sec:GTresults}

Our first result is that under logarithmic score, the game theoretic framework is essentially equivalent to the decision theoretic framework.

\begin{result} \label{res:GTlogscore}
The following sets are equal:
\begin{itemize}
\item Designs which are selected in SPEs of the game theoretic framework with logarithmic score and assumptions A1,A3.
\item Optimal designs in the decision theoretic framework with assumption A1 and logarithmic score.
\end{itemize}
\end{result}

The next result shows that for Hyv\"arinen score, the game theoretic framework addresses the issue of sensitivity to reparameterisation which occurred for the decision theoretic framework.
The result focuses on the case of a \emph{linear reparameterisation}
which we define as reparameterising $\theta$ to $B \theta$ for some invertible $B$.

\begin{result} \label{res:GTreparam}
The set of designs which are selected in SPEs of the game theoretic framework with Hyv\"arinen score and assumptions A1,A3 is unchanged by a linear reparameterisation.
\end{result}

Finally, our main result characterises which designs are selected in SPEs under Hyv\"arinen score.
It uses the following technical assumption:
\begin{enumerate}
\item[A4]
There exists some $\tau$ such that $\E_{\theta \sim \pi(\theta)}[\det \mathcal{I}(\theta; \tau)] > 0$.
\end{enumerate}
Informally this means there is a design under which the model is guaranteed to provide some information on every parameter, or linear combination of parameters.
We also use a regularity condition A5 (see Appendix \ref{app:scoring}).

\begin{result} \label{res:GTmain}
Consider the game theoretic framework with Hyv\"arinen score under assumptions A1, A3-A5.
Then the set of $\tau,A$ pairs which are selected in SPEs
are those which solve $\min_\tau \max_A \mathcal{K}(\tau,A)$
where
\begin{equation} \label{eq:K}
\mathcal{K}(\tau,A)
= -\E_{\theta \sim \pi(\theta)} \trace [ A^T \mathcal{I}(\theta; \tau) A ].
\end{equation}
Also, the set of designs which are selected in SPEs are those maximising
\begin{equation} \label{eq:Jadv}
\mathcal{J}_{\text{ADV}}(\tau) = \det \bar{\mathcal{I}}(\tau),
\end{equation}
where $\bar{\mathcal{I}}(\tau) = \E_{\theta \sim \pi(\theta)}[ \mathcal{I}(\theta; \tau) ]$.
\end{result}

\subsection{Adversarial objective properties} \label{sec:properties}

Using the decision theoretic framework with Hyv\"arinen score produced the objective
$\mathcal{J}_{\text{FIG}}(\tau) = \trace \bar{\mathcal{I}}(\tau)$.
Result \ref{res:GTmain} shows that the game theoretic framework instead produces the objective
$\mathcal{J}_{\text{ADV}}(\tau) = \det \bar{\mathcal{I}}(\tau)$.
This has improved properties, as we discuss in this subsection.

Result \ref{res:GTmain} also shows that maximisation of $\mathcal{J}_{\text{ADV}}(\tau)$ is equivalent to minimax optimisation of 
$\mathcal{K}(\tau,A)$, as defined in \eqref{eq:K}.
This is helpful in discussing the properties of $\mathcal{J}_{\text{ADV}}(\tau)$,
and particularly useful in defining a practical optimisation scheme,
as detailed below and in Section \ref{sec:opt}.

\paragraph{Linear reparameterisation invariance}

By Result \ref{res:GTreparam}, the set of optimal designs from $\mathcal{J}_{\text{ADV}}(\tau)$ is invariant to linear reparameterisation, unlike $\mathcal{J}_{\text{FIG}}(\tau)$.
This is also easy to show directly.
Consider a linear reparameterisation $\phi = B \theta$ with $\det B \neq 0$.
From \eqref{eq:reparam}, the FIM  is $\mathcal{I}_\phi(\phi; \tau) = B^{-T} \mathcal{I}_\theta(\theta; \tau) B^{-1}$.
Thus
$\bar{\mathcal{I}}_\phi(\tau) = B^{-T} \bar{\mathcal{I}}_\theta(\tau) B^{-1}$
and the reparameterised objective $\mathcal{J}_{\phi, \text{ADV}}(\tau)$ equals $\mathcal{J}_{\theta, \text{ADV}}(\tau) (\det B)^{-2}$:
the original objective multiplied by a positive constant.
Hence the set of optimal designs is unchanged.

\paragraph{Equivalent $A$ matrices}

From the cyclic property of trace,
\[
\mathcal{K}(\tau,A)
= -\E_{\theta \sim \pi(\theta)} \trace [ A A^T \mathcal{I}(\theta; \tau) ].
\]
Hence any two $A$ matrices producing the same $A A^T$ are equivalent
in that they give the same $\mathcal{K}(\cdot,A)$ function.
In particular, given any $A$, a Cholesky factor of $A A^T$ is equivalent.
So when we perform minimax optimisation, we can restrict $A$ to be a Cholesky matrix
(i.e.~lower triangular with positive diagonal entries)
with determinant 1.
This will help set up our optimisation algorithm later -- see Section \ref{sec:Arep}.

\paragraph{Intuitively informative designs}

As discussed in Section \ref{sec:FIGprop}, $\mathcal{J}_{\text{FIG}}(\tau)$ sometimes produces designs in which one parameter combination is inferred accurately but others are not.
The objective $\mathcal{J}_{\text{ADV}}(\tau)$ penalises such designs.
This is because it is the product of the eigenvalues of $\bar{\mathcal{I}}(\tau)$.
Thus, compared to $\mathcal{J}_{\text{FIG}}(\tau)$, it is much less advantageous to make one eigenvalue large and the others small.
See Figure \ref{fig:pk_posterior} for an illustration
-- the posterior under the ADV design is concentrated with respect to both parameters shown in this plot, unlike that under the FIG design.

A related property of the objective $\mathcal{J}_{\text{FIG}}(\tau)$ is that it sometimes produces singular or near-singular $\bar{\mathcal{I}}(\tau)$ matrices.
The objective $\mathcal{J}_{\text{ADV}}(\tau)$ avoids this by directly penalising matrices with low determinants.
One reason this is possible is that the determinant operation is affected by off-diagonal elements of $\bar{\mathcal{I}}(\tau)$, unlike the trace.

\paragraph{Bayesian justification}

Unlike $\mathcal{J}_{\text{FIG}}(\tau)$, $\mathcal{J}_{\text{ADV}}(\tau)$ is not defined as the expectation of a utility function.
Therefore the ``pseudo-Bayesian'' or ``fully Bayesian'' definitions of \cite{Ryan:2016} discussed in Section \ref{sec:utilities} cannot be directly applied.
However, $\mathcal{J}_{\text{ADV}}(\tau)$ meets the spirit of the pseudo-Bayesian definition as it depends on FIMs rather than posteriors.
Nonetheless, we have shown it emerges from a game theoretic framework based on enabling an experimenter to estimate the posterior well.

\paragraph{Link to $D$-optimality}

Classical optimal design using $D$-optimality requires a design to maximise $\det \mathcal{I}(\theta; \tau)$ at a reference $\theta$ value.
This can be generalised to a Bayesian setting incorporating parameter uncertainty in several ways.
See \cite{Atkinson:2007}, Table 18.1, for a list of 5 possible objectives, including $\mathcal{J}_{\text{ADV}}(\tau)$
(and the objective of \cite{Pronzato:1985}, mentioned in Section \ref{sec:related}, which differs from $\mathcal{J}_{\text{ADV}}(\tau)$ by swapping the order of determinant and expectation.)
One contribution of our work is to provide theoretical support and an optimisation method for this particular choice.

\paragraph{Computational tractability}

It is hard to optimise $\mathcal{J}_{\text{ADV}}(\tau)$ directly for continuous $\tau$, as it is not easy to obtain an unbiased gradient estimate due to the non-linearity of the determinant operator.
However optimisation based on \eqref{eq:K} is tractable, as described in the next section.

\section{Optimisation} \label{sec:opt}

Result \ref{res:GTmain} of the previous section
motivates performing optimal design by finding minimax solutions of
$\mathcal{K}(\tau,A) = -\E_{\theta \sim \pi(\theta)} \trace [ A^T \mathcal{I}(\theta; \tau) A ]$.
Since this is an expectation,
computation of unbiased gradient estimates is straightforward
when $\mathcal{I}(\theta; \tau)$ can be evaluated.
(See Appendix \ref{sec:estimationMore} for discussion of the case where this is not possible.)
This section describes how as a consequence it is possible to find candidate minimax solutions using generic gradient based optimisation methods.
Throughout we assume $\nabla_\tau \mathcal{I}(\theta; \tau)$ and $\nabla_\tau \E_{\theta \sim \pi(\theta)} \mathcal{I}(\theta; \tau)$ exist.

Section \ref{sec:GDA} gives background on gradient based optimisation methods for minimax problems.
Section \ref{sec:Arep} discusses how to deal with the constraint $\det A = 1$.
Section \ref{sec:Kgrad} describes calculation of unbiased gradient estimates.
Section \ref{sec:main alg} presents our optimisation algorithm,
and describes various implementation details.

\subsection{Gradient descent ascent} \label{sec:GDA}

Algorithm \ref{alg:GDA}, \emph{gradient descent ascent}, attempts to solve $\min_x \max_y f(x,y)$.
It iteratively updates $(x_t, y_t)$ based on $g_{x,t}$ and $g_{y,t}$,
unbiased gradient estimates of $-\nabla_x f(x_t,y_t)$ and $\nabla_y f(x_t,y_t)$.
For an overview and history of GDA see \cite{Lin:2019}.
GDA generalises \emph{stochastic gradient descent} (SGD),
which is the special case where $f$ is a function of $x$ alone,
and only $x_t$ updates are needed.

\begin{algorithm}[htbp] \caption{Gradient descent ascent (GDA)} \label{alg:GDA}
\begin{algorithmic}[1]
\STATE Input: Initial values $x_1, y_1$, update subroutines $h_x, h_y$.
\FOR{$t = 1,2,\ldots$}
\STATE Compute $g_{x,t}, g_{y,t}$, unbiased estimates of $-\nabla_x f(x_t,y_t)$ and $\nabla_y f(x_t,y_t)$.
\STATE Update estimates using $x_{t+1} = x_t + h_x(g_{x,t})$, $y_{t+1} = y_t + h_y(g_{y,t})$.
\ENDFOR
\end{algorithmic}
\end{algorithm}

A simple update rule for GDA,
which we will refer to as the \emph{default} update rule,
is $z_{t+1} = z_t + \alpha_{z,t} g_{z,t}$ (for $z \in \{ x,y \}$),
given predefined learning rate sequences $\alpha_{x,t}, \alpha_{y,t}$.
Stochastic approximation theory \citep[see e.g.][]{Kushner:2003} suggests using learning rates such that $\sum_{t=1}^\infty \alpha_{z,t} = \infty$,
$\sum_{t=1}^\infty \alpha_{z,t}^2 < \infty$
(for $z \in \{x,y\}$).
This ensures that SGD using this update rule and unbiased gradient estimates converges to a local minimum, under appropriate regularity conditions.

More sophisticated update rules have been developed,
effectively tuning the learning rates adaptively and using different learning rates for each component of the $x$ and $y$ vectors.
We use the popular Adam update rule \citep{Kingma:2015}.
(We use the default Adam tuning parameters,
which typically produce good empirical performance but not asymptotic convergence.
Convergence guarantees are possible by setting some parameters to decay: see \citealp{Kingma:2015}.)

Convergence of GDA under any update rule is more complicated than SGD,
and is an area of active research.
One issue is that the dynamics can produce limit cycles as well as limit points.
Algorithms to avoid this have been suggested, including a \emph{two time-scale update rule} \citep{Borkar:1997, Heusel:2017} in which $\lim_{t \to \infty} \alpha_{y,t} / \alpha_{x,t} = \infty$
(or a similar condition under the Adam learning rule.)
Another issue is to characterise the limit points of GDA as an appropriate local generalisation of minimax solutions \citep{Heusel:2017, Jin:2019, Lin:2019}.

However, for our experimental design application, we find empirically that standard GDA methods with the Adam update rule suffice to produce sensible designs.
Therefore we recommend using GDA with this update rule,
and checking for convergence using diagnostics, multiple runs from different initial values, and post-processing.
Details of these are contained in the following subsections.

\subsection{Representation of $A$} \label{sec:Arep}

We wish to solve $\min_\tau \max_A \mathcal{K}(\tau,A)$
under the constraint that $\det A = 1$.
As discussed in Section \ref{sec:properties}, it is sufficient to search for $A$ over Cholesky matrices
(i.e.~lower triangular with positive diagonal entries)
with determinant 1.
Such matrices can be represented as
\begin{equation} \label{eq:A(eta)}
A(\eta) =
\begin{pmatrix}
\exp(\eta_{11}) & 0           & \ldots & 0 \\
\eta_{21}       & \exp(\eta_{22}) & \ldots & 0 \\
\vdots      & \vdots      & \ddots & \vdots \\
\eta_{p1}       & \eta_{p2}       & \ldots & \exp(-\sum_{i=1}^{p-1} \eta_{ii})
\end{pmatrix}.
\end{equation}
This maps an unconstrained real vector $\eta$ of $\eta_{ij}$ variables to the space of matrices of interest.
We can now solve $\min_\tau \max_\eta \mathcal{K}(\tau,A(\eta))$ using GDA.
We initialise $\eta = 0$ so that initially $A=I$, corresponding to the critic making no reparameterisation.

\subsection{Gradient estimation} \label{sec:Kgrad}

In this section we will consider estimating $\nabla_z \mathcal{K}(\tau, A(\eta))$ for $z \in \{\tau, \eta\}$
in the common case where it is easy to evaluate the FIM and related gradients.
See Appendix \ref{sec:estimationMore} for discussion of cases where these are intractable.

In a few cases it is possible to directly evaluate 
$\bar{\mathcal{I}}(\tau) = \E_{\theta \sim \pi(\theta)}[ \mathcal{I}(\theta; \tau) ]$.
See Sections \ref{sec:illustration} and \ref{sec:geostat} for example.
Then, using \eqref{eq:K},
we have $\mathcal{K}(\tau, A) = - \trace[ A^T \bar{\mathcal{I}}(\tau) A ]$,
and gradients can be evaluated using automatic differentiation \citep{Baydin:2017} or derivation of direct expressions for them.
Our code performs automatic differentiation using the PyTorch library \citep{Paszke:2019}.

More commonly it is necessary to derive unbiased gradient estimates.
From the definition of $\mathcal{K}(\tau, A)$, \eqref{eq:K},
an unbiased estimate is
\begin{equation} \label{eq:Kest}
\hat{\mathcal{K}}(\tau, A) = -\trace \left[A^T \left\{ \frac{1}{K} \sum_{k=1}^K \mathcal{I}(\theta^{(k)}; \tau) \right\} A \right].
\end{equation}
We also have the following result.
\begin{result} \label{res:unbiased}
Under appropriate regularity conditions on $\trace [A^T \mathcal{I}(\theta; \tau) A]$
(see Appendix \ref{sec:regularity}),
then $\nabla_z \hat{\mathcal{K}}(\tau, A(\eta))$ is an unbiased estimate of
$\nabla_z \mathcal{K}(\tau, A(\eta))$ for $z \in \{\tau, \eta\}$.
\end{result}

We calculate gradients of $\hat{\mathcal{K}}(\tau, A)$ using automatic differentiation in PyTorch.
Note that using a larger $K$ in \eqref{eq:Kest}
reduces the variance but increases computational cost.
We explore this trade-off in Section \ref{sec:pk}.

\subsection{Main algorithm} \label{sec:main alg}

Algorithm \ref{alg:GDA_BED} applies GDA to our experimental design setting.
Note that we perform $R$ replications of GDA in parallel from different initial conditions.
This is typically more efficient than repeating the algorithm $R$ times in serial.
One reason is that the calculations in step 3 and 4 are amenable to parallelisation.
(We ran our experiments on a CPU, but our PyTorch code can easily exploit GPU parallelisation, allowing for further speed improvements.)
Another reason is that the same simulations in step 3 can be reused for all replications.
Algorithm \ref{alg:GDA_BED} can also be used for SGD to optimise the FIG objective by keeping $\eta=0$ and only updating $\tau$.

\begin{algorithm}[htbp] \caption{Gradient descent ascent for Bayesian experimental design} \label{alg:GDA_BED}
\begin{algorithmic}[1]
\STATE Input:
Number of samples to use in \eqref{eq:Kest} $K$,
number of parallel replications to perform $R$,
initial values $\tau_1^i, \eta^i_1$ (for $i = 1,2,\ldots,R$),
update subroutines $h_\tau, h_\eta$.
\FOR{$t = 1,2,\ldots$}
\STATE Sample $\theta^{(k)}$ from the prior for $k = 1, 2, \ldots, K$.
\STATE Compute $g^i_{\tau,t}, g^i_{\eta,t}$, unbiased estimates of $-\nabla_\tau \mathcal{K}(\tau_t^i, A(\eta_t^i))$ and $\nabla_\eta \mathcal{K}(\tau_t^i, A(\eta_t^i))$ using automatic differentiation\footnotemark of \eqref{eq:Kest} for all $i$.
\STATE Update estimates using $\tau^i_{t+1} = \tau^i_t + h_\tau(g^i_{\tau,t})$, $\eta^i_{t+1} = \eta^i_t + h_\eta(g^i_{\eta,t})$ for all $i$.
\ENDFOR
\end{algorithmic}
\end{algorithm}
\footnotetext{
In practice we implement this by differentiating
$\sum_{i=1}^R \hat{\mathcal{K}}(\tau_t^i, A(\eta_t^i))$
with respect to all $\tau, \eta$ variables.
This is more efficient in backwards mode automatic differentiation, as implemented in PyTorch.}

In the remainder of this section various implementation details are discussed.

\subsubsection{Diagnostics}

Step 4 of Algorithm \ref{alg:GDA_BED} involves calculating $\hat{\mathcal{K}}(\tau,A)$ from \eqref{eq:Kest},
whose gradients are then found using automatic differentiation.
This Monte Carlo estimate of $\mathcal{K}$ can be used as a diagnostic of the algorithm's performance.
As we are performing minimax optimisation, it will typically rise and fall before reaching an equilibrium.
Also the values of $\hat{\mathcal{K}}$ from different parallel runs can be highly correlated since they are based on the same $\theta^{(k)}$ samples.
Both phenomena can be seen in the bottom left graph of Figure \ref{fig:pk_GDA_traces}.
(The presence of correlation is illustrated by the fact that the right hand side of the graph seems to show a single thick line.
In fact there are multiple lines with different values which are highly correlated with each other.)

An alternative diagnostic is to estimate $\mathcal{J}_{\text{ADV}}$ as defined in \eqref{eq:Jadv} by
\begin{equation} \label{eq:Jhat}
\hat{\mathcal{J}}_{\text{ADV}}(\tau) = \det \left( \frac{1}{J} \sum_{j=1}^J \mathcal{I}(\tilde{\theta}^{(j)}; \tau) \right).
\end{equation}
Our code can optionally calculate $\hat{\mathcal{J}}_{\text{ADV}}$ for designs produced during its execution.
To do so we initially sample $\tilde{\theta}^{(j)}$ values from the prior for $j = 1,2,\ldots,J$ (we take $J=1000$).
Then $\hat{\mathcal{J}}_{\text{ADV}}$ is calculated for each design $\tau_t^i$ produced during the algorithm.
Unlike $\hat{\mathcal{K}}$, this adds an extra computational cost.
However this diagnostic has the advantage that it directly estimates the objective $\mathcal{J}_{\text{ADV}}$ so larger values correspond to better performing designs.
For example, the bottom right graph of Figure \ref{fig:pk_GDA_traces} directly traces the performance of designs during optimisation.

For SGD optimisation we can return $\hat{\mathcal{K}}(\tau,I)$ as a diagnostic.
This directly estimates $-\mathcal{J}_{\text{FIG}}(\tau)$ but does so using a different $\theta$ sample each time, which adds some variability to the diagnostic.
To remedy this we can also calculate
\[
\hat{\mathcal{J}}_{\text{FIG}}(\tau) = \trace \left( \frac{1}{J} \sum_{j=1}^J \mathcal{I}(\tilde{\theta}^{(j)}; \tau) \right)
\]
for a fixed sample of $\tilde{\theta}^{(j)}$ values, as above.

\subsubsection{Termination}

We run Algorithm \ref{alg:GDA_BED} for a fixed number of iterations or fixed runtime.
Alternatively it could be run until a convergence condition is met for one of the diagnostics above,
or for the size of the increments to $\tau$ or $A$.

\subsubsection{Optimisation under constraints}

We often wish to find the optimal design under a constraint: $\tau \in \mathcal{T} \subset \mathbb{R}^d$.
In our applications we achieved this using simple pragmatic approaches.
In Section \ref{sec:illustration} we represent (scalar) $\tau$ as the transformation of an unconstrained variable.
For most examples in Section \ref{sec:pk} the designs remain in $\mathcal{T}$ under unconstrained optimisation, so no modification to this is needed.
We address constraints for analyses that require a minimum time between observations in Section \ref{sec:pk} or in Section \ref{sec:geostat} by adding a large penalty to $\mathcal{K}$ for $\tau \not \in \mathcal{T}$, whose gradient moves designs towards $\mathcal{T}$.
In more complex settings these methods may not suffice.
A more sophisticated alternative would be to compose each GDA update with a projection operation into $\mathcal{T}$ \citep{Kushner:2003}.

\subsubsection{Local optima}

GDA can often converge to multiple possible locally optimal designs.
To attempt to find the global optimum we run GDA multiple times from different initial values of $\tau$ (keeping $\eta$ initialised as a zero vector),
and compare their $\hat{\mathcal{J}}$ diagnostics.

In some settings we can also use a \emph{point exchange algorithm}.
Suppose we must select multiple \emph{design points} from some region.
Optimal designs often have a high degree of \emph{replication}: they consist of a small number of clusters of repeated observations \citep{Gotwalt:2009, Overstall:2017, Binois:2019}.
While gradient based optimisation can find the optimal cluster locations well,
there may be a large number of local optima, differing by the number of points in each cluster.
See Figure \ref{fig:pk_GDA_designs} for an example.
A point exchange algorithm takes cluster locations as input and uses discrete optimisation to find optimal cluster sizes.
We use a simple approach detailed in Appendix \ref{sec:pointExchange}.
There is scope for developing more sophisticated approaches in future work.

\section{Poisson model illustration} \label{sec:illustration}

This section provides a simple illustrative example of the properties of the SIG, FIG and ADV approaches to experimental design.
The setting is that an experimenter must divide a unit of time between two experiments making Poisson observations with different rates.
More precisely, the design is $\tau \in [0,1]$.
There are 2 independent observations $y_1 \sim Poisson(\tau \theta_1 \omega_1),
y_2 \sim Poisson([1-\tau] \theta_2 \omega_2)$.
We assume $\omega_1 > \omega_2 > 0$ and that $\theta_1, \theta_2$ have independent $Gamma(2,1)$ priors.

\paragraph{Analytic results}

Optimal designs for this example can be derived analytically.
We summarise the results here: see Appendix \ref{app:Poisson} for derivations.
The optimal design is $\tau=1$ under FIG and $\tau=1/2$ under SIG or ADV.
Under the FIG design, $y_2$ is always zero, so this design produces no information on $\theta_2$
i.e.~the posterior always equals the prior.
The SIG/ADV design avoids this undesirable property.
Also, the SIG/ADV design is invariant to linear reparameterisations.
However for FIG, linear reparameterisation can change the optimal design to $\tau_1=0$, or make all values of $\tau$ optimal.

\paragraph{Numerical optimisation}

We perform a numerical analysis of this example with $\omega_1=2, \omega_2=1$.
Since the design $\tau \in [0,1]$ is bounded
we optimise a transformation, $\lambda = \text{logit}(\tau)$,
to allow the use of unconstrained optimisation.
Following \eqref{eq:A(eta)} we take the critic's action to be
\[
A(\eta) =
\begin{pmatrix}
\exp(\eta_{11}) & 0           \\
\eta_{21}       & \exp(-\eta_{11})
\end{pmatrix}.
\]

\begin{figure}[htp]
\begin{center}
\includegraphics[width=0.7\textwidth]{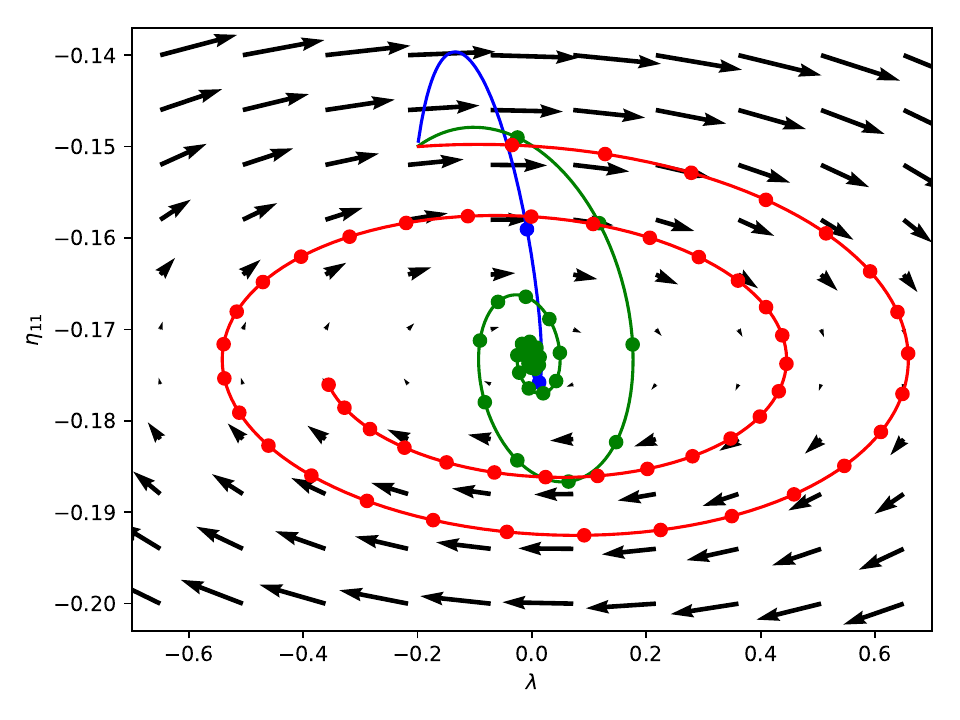}
\end{center}
\vspace{-5mm}
\caption{GDA vector field and trajectories for the Poisson example.
Paths are shown for experimenter learning rate $10^{-2}$
and critic learning rate $10^{-3}$ (blue), $10^{-4}$ (green), $10^{-5}$ (red).
Points indicate every 500 steps of optimisation.
The vector field corresponds to the $10^{-5}$ critic learning rate.}
\label{fig:illustration}
\end{figure}

Figure \ref{fig:illustration} shows a GDA vector field using default update rules with a particular choice of learning rates.
The figure shows $\lambda$ and $\eta_{11}$ for $\eta_{21}$ fixed at zero.
(In practice other $\eta_{21}$ values quickly converge to zero.)
The vector field illustrates spiral trajectories converging to a limit point.
Additionally, it shows that for any fixed value of $\eta_{11}$, $\frac{\partial \mathcal{K}}{\partial \lambda}$ has a fixed sign.
This illustrates that for fixed $A$ (i.e.~the FIG setting),
SGD optimisation produces a design converging to either $\tau=0$ or $1$.

The figure also shows GDA trajectories as the critic learning rate is varied.
In all cases, the trajectories converge on the limit point.
However, convergence is much faster as the critic's learning rate is increased.

\section{Pharmacokinetic example} \label{sec:pk}

This section contains simulations studies on a pharmacokinetic model.
The main goal is to investigate the performance of our FIG and ADV approaches and compare them with existing methods for SIG.

\subsection{Model}

Pharmacokinetics studies the time course of drug absorption,
distribution, metabolism, and excretion.
Concentration level is observed via samples of fluid
-- such as blood, plasma or urine --
from the subject taken at preplanned time points.
The number of observations is constrained by budget and resources,
as well as patient comfort and wellbeing.
We assume observed concentration, $y_i$, at time $\tau_i$ (in hours) is distributed as
\[
y_i \sim N(x(\theta, \tau_i), \sigma^2),
\qquad \text{where}\ 
x(\theta, \tau_i) = \frac{D\theta_2 (\exp[-\theta_1 \tau_i] - \exp[-\theta_2 \tau_i]),}{\theta_3(\theta_2 - \theta_1)},
\]
and $D=400$.
Concentrations at different times are assumed to be independent.
We assume independent log normal priors
\[
\theta_1 \sim LN(\log 0.1, 0.05), \theta_2 \sim LN(\log 1, 0.05), \theta_3 \sim LN(\log 20, 0.05),
\]
and aim to find $15$ observation times in $[0,24]$. Also we treat $\sigma^2 = 0.1$ as known.

A similar model to the above was used by \cite{Ryan:2014} and \cite{Overstall:2017}.
However we make two modifications to create a simple setting for comparing different methods.
First, we omit a multiplicative noise term.
Secondly, we do not enforce a 15 minute minimum time gap between consecutive observations, as its implementation would vary between methods making it more difficult to draw fair conclusions.
In Section \ref{sec:pk_realistic} we remove these modifications and show results from our approach for the realistic model used in previous work.

\subsection{Methods} \label{sec:pk_methods}

The FIM for this model is available in closed form --
see Appendix \ref{app:pkFIG}.
This allows the optimal ADV design to be found using GDA (Algorithm \ref{alg:GDA_BED}),
and the optimal FIG design using a SGD variant (only updating $\tau$).
In both cases we used the Adam update rule.
When implementing these algorithms, we found no need to use constrained optimisation as the designs remained in the interval $[0,24]$ in any case.
We performed 100 replications in parallel with each algorithm, based on 100 initial designs sampled from a uniform distribution over $[0,24]^{15}$.
We considered several choices of $K$, the number samples to estimate $\mathcal{K}$ in \eqref{eq:Kest}, and selected $K=1$.
See Appendix \ref{app:Kcomparison} for details.

We compare our results to two methods of finding the optimal SIG design:
the approximate coordinate exchange (ACE) algorithm of \cite{Overstall:2017},
and the prior contrastive estimation (PCE) algorithm of \cite{Foster:2020}.
(PCE is one of several methods in \citealt{Foster:2020}, and is not their overall recommendation. However we found PCE converged more quickly than the alternatives for this example, so we use it as lower bound on the speed of their methods.)
Implementation details of these methods are given in Appendices \ref{app:pkACE} and \ref{app:pkPCE}.
PCE allowed 100 replications to be run in parallel.
We ran replications of ACE serially, noting that the ACE code utilises multiple cores during its execution in any case. 
As ACE took much longer to run we used only 30 replications.

\subsection{Results}

Table \ref{tab:timings} shows mean times for each method
i.e.~run time divided by number of designs returned,
demonstrating the speed advantage of GDA and SGD.
Below we comment in more detail on each algorithm's results.

\begin{table}[hbtp]
\begin{tabular}{c|ccccccc}
                      & GDA & GDA+PE & SGD & SGD+PE & ACE  & PCE \\
\hline
Mean time (seconds)   & 1.4 & 2.2    & 1.4 & 1.5    & 8012 & 36  \\
Number of repetitions & 100 & 100    & 100 & 100    & 30   & 100
\end{tabular}
\caption{Mean times to run optimisation methods on the pharmacokinetic example.
(n.b.~PE is ``point exchange'').
Both ACE and PCE have not fully converged after the time quoted, so the figures given are lower bounds on the run time.
}
\label{tab:timings}
\end{table}

\paragraph{ADV}

\begin{figure}[htp]
\begin{center}
\includegraphics[width=0.45\textwidth]{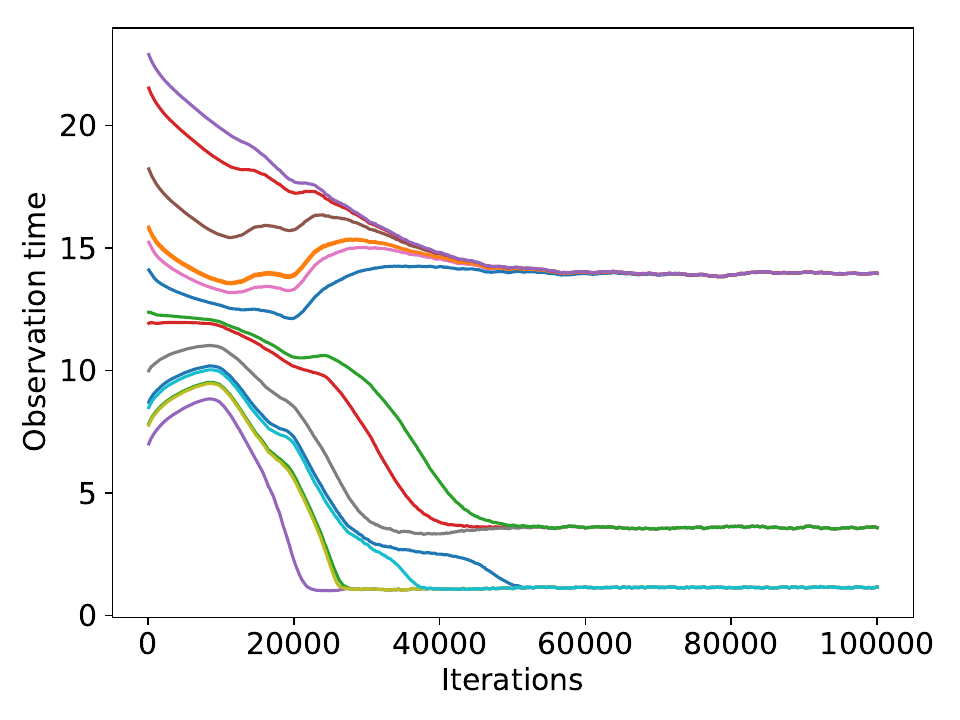}
\includegraphics[width=0.45\textwidth]{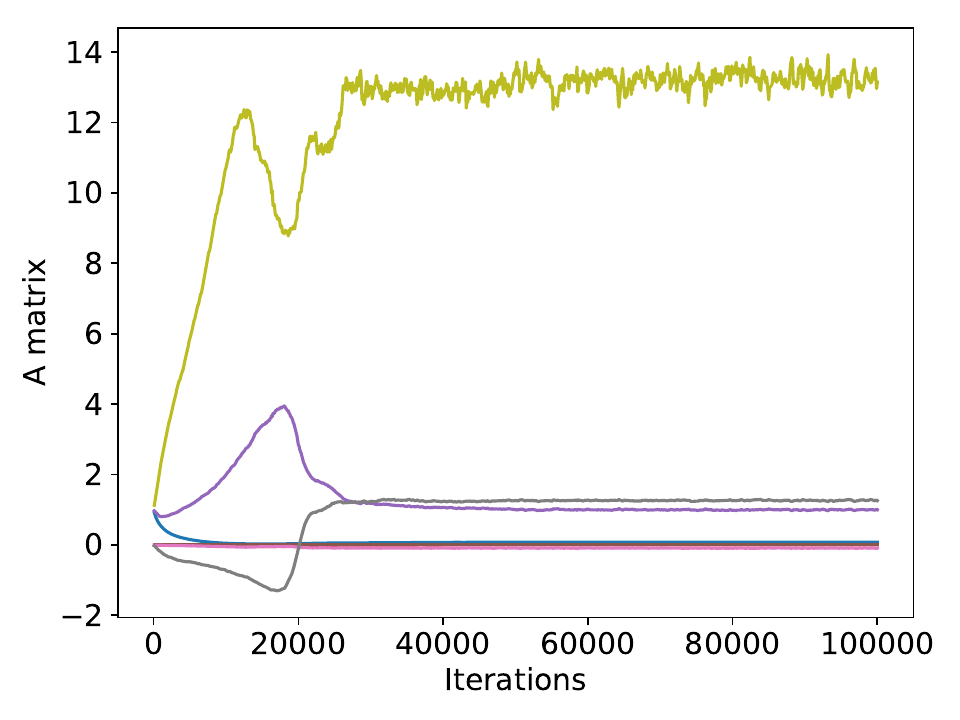} \\
\includegraphics[width=0.45\textwidth]{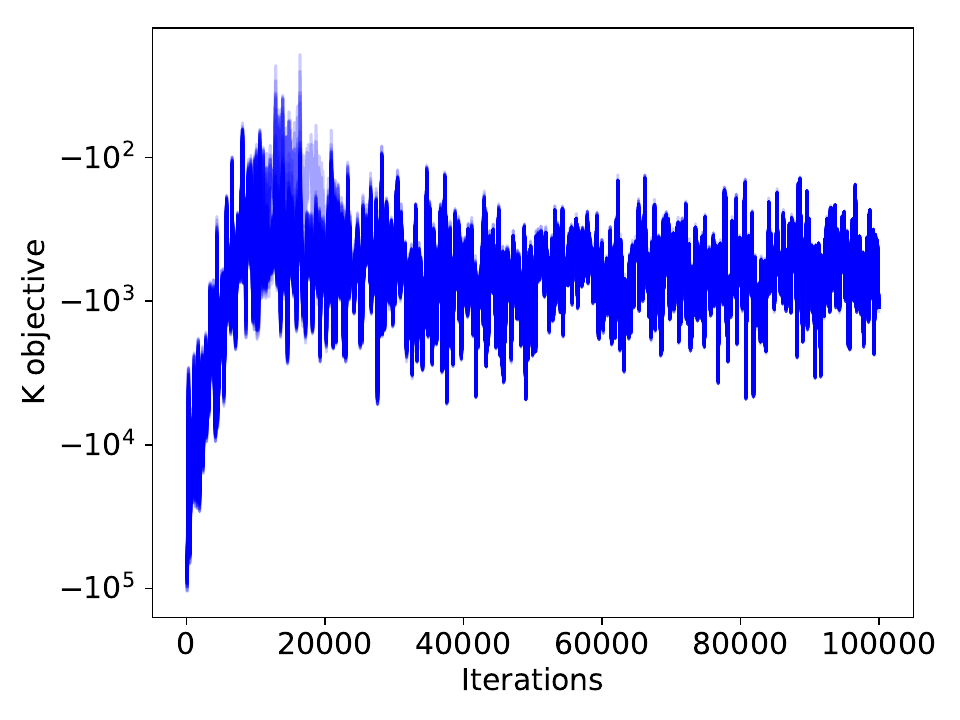}
\includegraphics[width=0.45\textwidth]{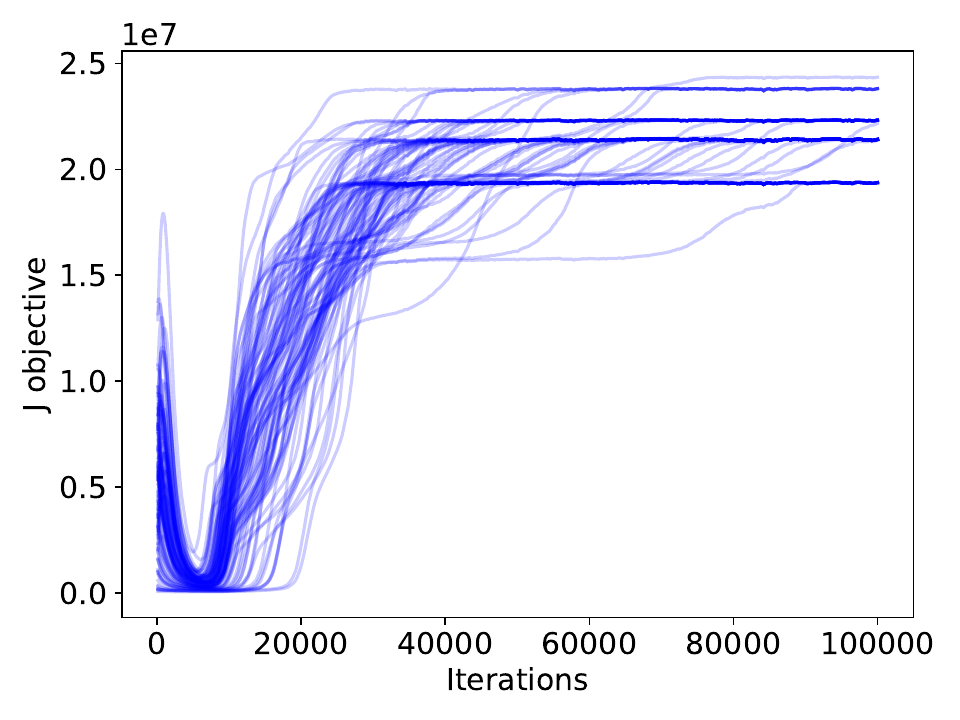}
\end{center}
\caption{Trace plots for GDA on the pharmacokinetic example.
The top row shows output for a single run of GDA.
The bottom row shows output for 100 runs.}
\label{fig:pk_GDA_traces}
\end{figure}

The top half of Figure \ref{fig:pk_GDA_traces} shows trajectories of $\tau$ and $A$ during a single replication of GDA optimisation.
The design points eventually settle into clusters of repeated observations at three times.
The bottom half shows estimated $\mathcal{K}$ and $\mathcal{J}$ objectives over 100 replications.
Although all runs have converged after $100,000$ iterations, the objective values are not identical.

\begin{figure}[htp]
\begin{center}
\includegraphics[width=0.45\textwidth]{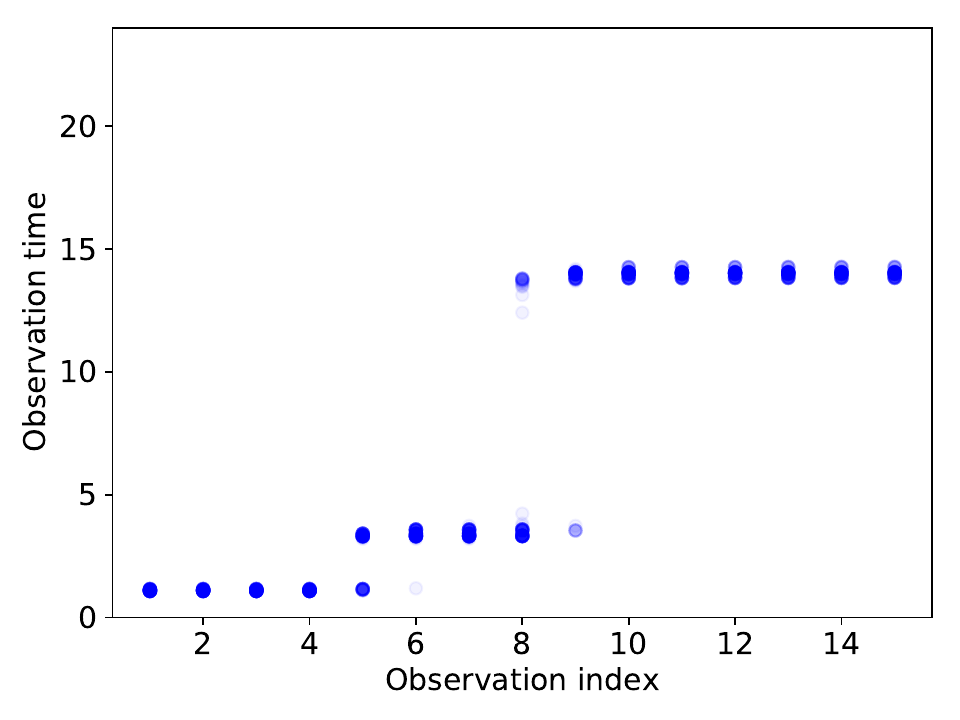}
\includegraphics[width=0.45\textwidth]{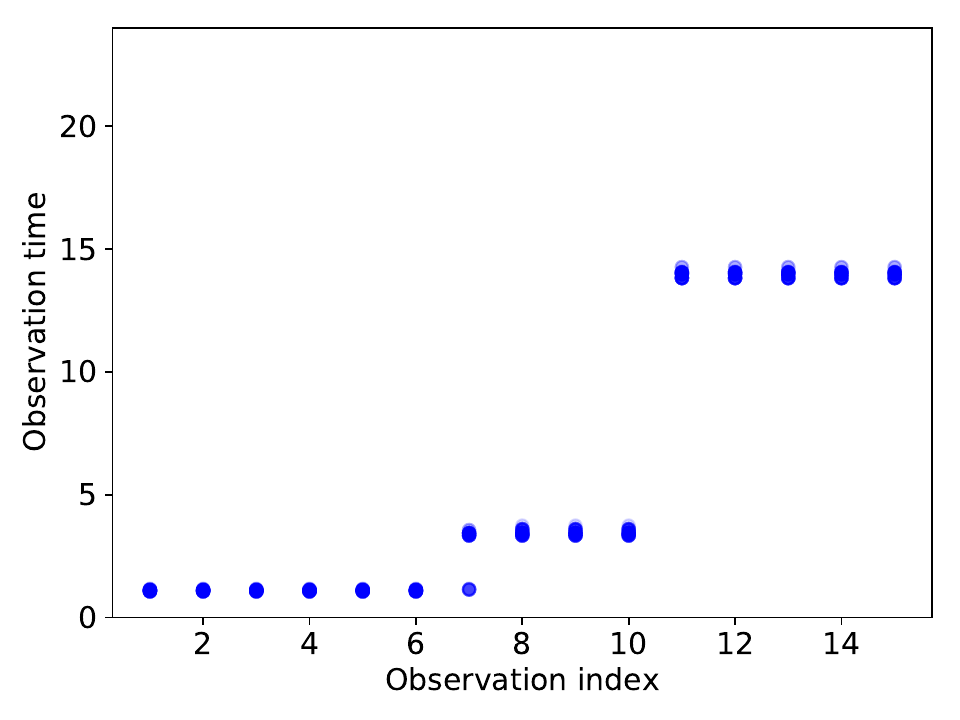} \\
\end{center}
\caption{Designs output from 100 optimisation runs for the pharmacokinetic example.
The horizontal axis shows the index of each point in the sorted design
i.e.~observation times are shown in increasing order from left to right.
Points are plotted as semi-transparent, so rare results are light, and common results are dark.
The left plot shows GDA output after 100,000 iterations.
In the right plot point exchange is also applied.
}
\label{fig:pk_GDA_designs}
\end{figure}

Figure \ref{fig:pk_GDA_designs} shows final designs for all replications.
The design points typically converge to 3 clusters,
around times $1.1$, $3.4$ and $14.0$.
However the cluster sizes vary between runs, as different local optima are found.
This explains why runs converge to slightly different objective values.
Employing point exchange reduces the variation in cluster sizes.
(There is some variability in exact cluster locations.
Further investigation showed that this is mainly due to cluster locations changing slightly depending on cluster sizes.)
PE finds two candidates for an optimal design, with 6 or 7 points near time 1.
(To find an overall optimum one could now estimate $\hat{\mathcal{J}}_{\text{ADV}}$ for both using a larger sample size.)
However GDA runs produced a maximum of 6 points for this cluster.
This highlights the importance of PE to find global optima.

\begin{figure}[htp]
\begin{center}
\includegraphics[width=0.45\textwidth]{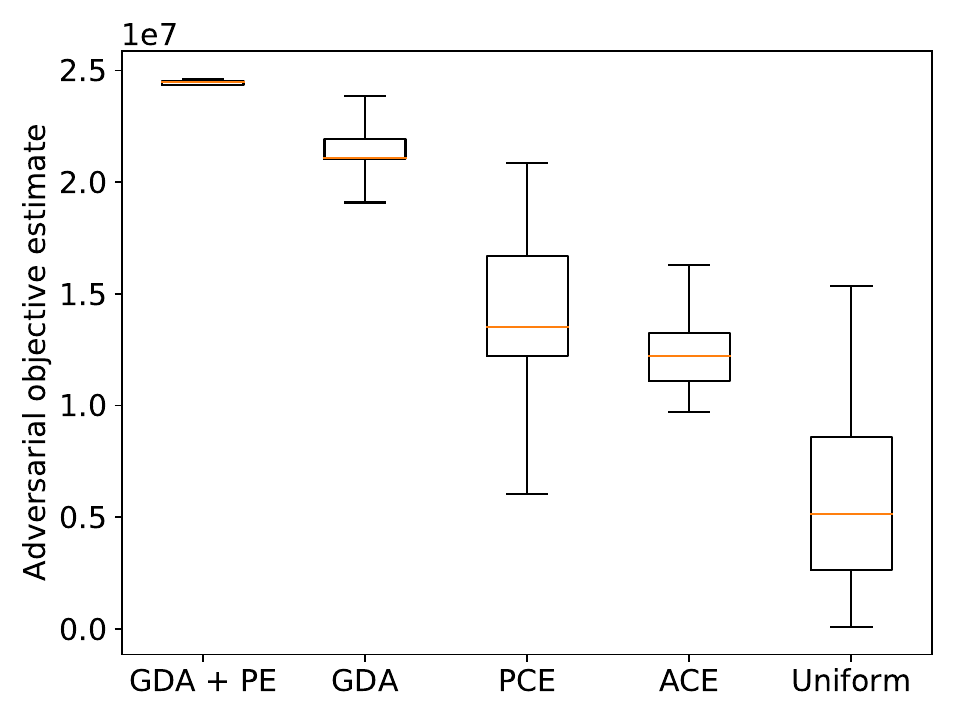}
\includegraphics[width=0.45\textwidth]{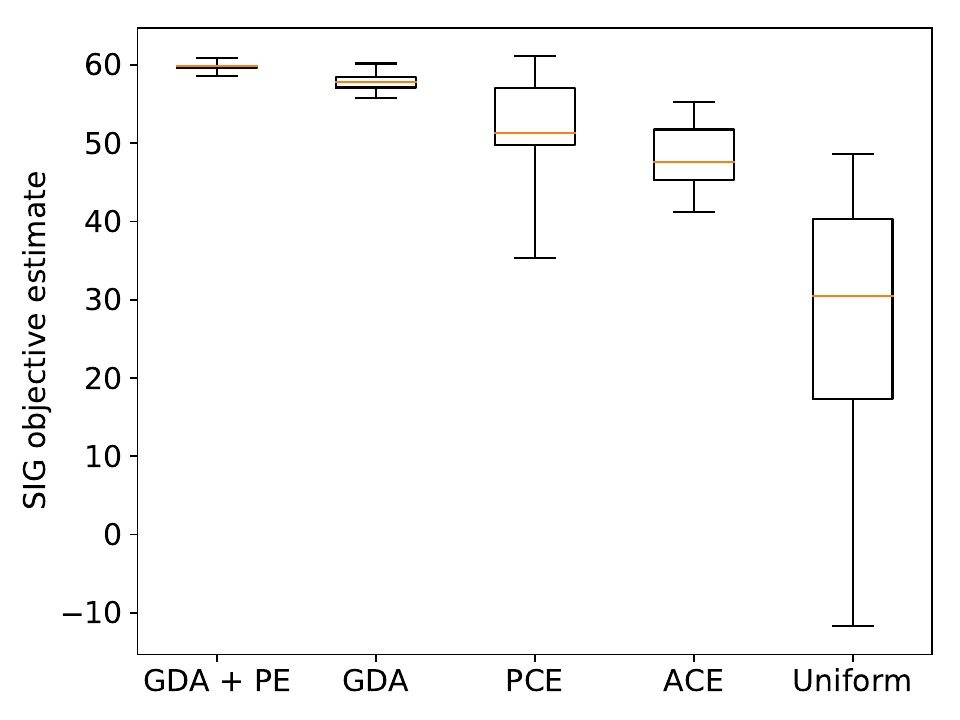}
\end{center}
\caption{Performance of pharmacokinetic example designs output by various methods.
Performance is judged by estimated objective value achieved for (left) our adversarial approach and (right) the Shannon information gain approach.
Boxplots are shown for estimates taken from 100 designs (or 30 in the case of ACE) from repeated optimisation runs with different initialisations.}
\label{fig:pk_boxplots}
\end{figure}

Figure \ref{fig:pk_boxplots} (left) shows $\hat{\mathcal{J}}_{\text{ADV}}$ values achieved by the final designs from GDA and the SIG methods, as well as for designs sampled from a uniform distribution over $[0,24]^{15}$.
These estimated were calculated using equation \eqref{eq:Jhat}.
On average GDA outperforms the SIG methods
and further improvement is achieved by also applying point exchange.

\begin{figure}[htp]
\begin{center}
\includegraphics[width=0.7\textwidth]{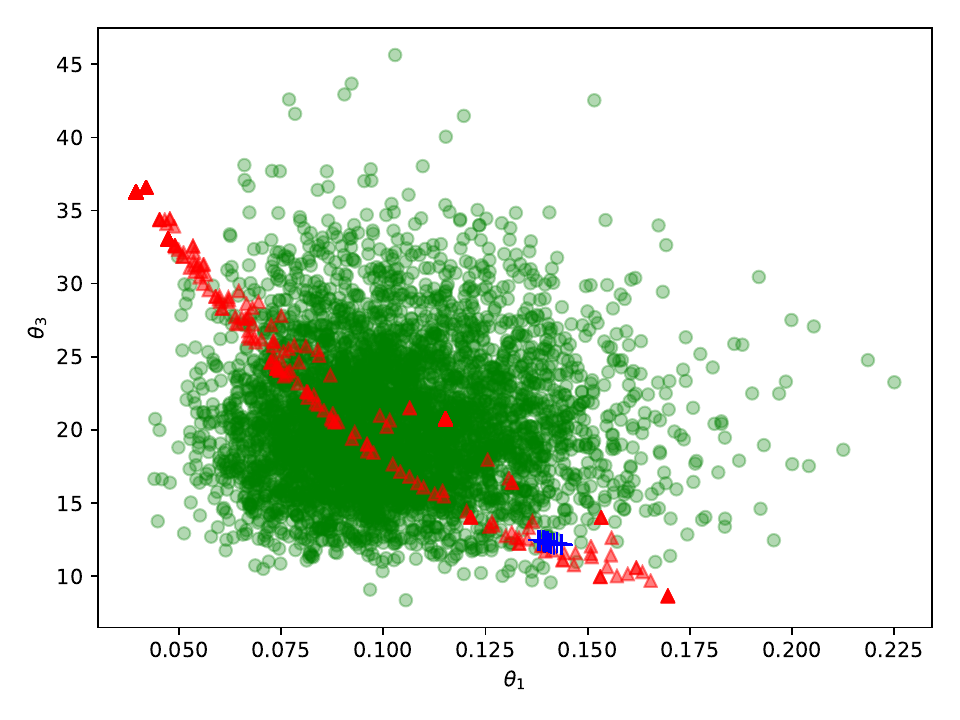}
\end{center}
\caption{Bivariate posteriors for $\theta_1, \theta_3$ in the pharmacokinetic example.
Samples are plotted from: the prior (green circles),
the posterior with a FIG design (red triangles),
the posterior with a ADV design (blue pluses).
The ADV and FIG designs used 100,000 iterations of GDA or SGD,
followed by point exchange optimisation.
The true parameter values, sampled from the prior, are $\theta_1=0.14,  \theta_3=12.29$, consistent with the blue pluses.
Datasets for use in inference were then sampled from the model,
using the same observation noise realisations for both ADV and FIG designs.
Posterior samples are importance sampling output.}
\label{fig:pk_posterior}
\end{figure}

Finally, Figure \ref{fig:pk_posterior} illustrates a posterior found from one particular ADV design, and shows that it is concentrated around the true parameters.

\paragraph{FIG}

Full details of SGD optimisation results are given in Appendix \ref{app:pkSGD}.
Briefly, SGD designs always converge to a single cluster around time 12.
Intuitively this is a poor design:
the three $\theta$ parameters of the $x(\theta,\tau)$ function cannot all be identified from repeated observations at a single time point $\tau$.
Figure \ref{fig:pk_posterior} illustrates that this design indeed performs poorly by presenting a typical posterior.
The narrowness of the FIG design posterior illustrates that the posterior is highly concentrated for some function of $\theta_1$ and $\theta_3$.
However the posterior is diffuse for $\theta_1$ and $\theta_3$ marginally:
it stretches almost the full length of the prior distribution,
unlike the highly concentrated ADV design posterior.

\paragraph{SIG}

Running ACE under the default tuning choices took a few hours, much slower than GDA which took only a few minutes to run 100 repetitions.
Furthermore, the ACE results do not seem to have converged to the optimal design in the time.
See Appendix \ref{app:pkACE} for more details.
(We also explored varying one tuning choice in ACE: the number of Monte Carlo samples for utility estimates used to fit the Gaussian process.
However we found only marginal improvements over the recommended default tuning for ACE.)

We ran PCE for roughly 10 times longer than GDA and found the results concentrated near 3 observation times similar to those for GDA.
However this runtime appears to only be a lower bound on the time needed for convergence, as running PCE for longer increased the concentration of the design points.
We also explored alternative methods from \cite{Foster:2020} but were not able to improve on the PCE results.
See Appendix \ref{app:pkPCE} for more details.

Figure \ref{fig:pk_boxplots} (right) shows estimates of expected Shannon information gain achieved by the final designs from all SIG methods, as well as GDA designs and designs sampled from a uniform distribution over $[0,24]^{15}$.
The calculation method is described in Appendix \ref{app:pkSIG}.
On average GDA outperforms the other methods, with a slight further improvement from using point exchange.
This suggests that GDA designs produce good performance under the SIG objective,
and is also further evidence that the ACE and PCE methods have not fully converged to the overall optimum.

\subsection{Conclusion of comparisons}

We have shown that our optimisation method to find ADV and FIG designs is faster than SIG optimisation methods by at least a factor of 10.
The true advantage may be greater as the SIG methods do not appear to have fully converged in the time stated.

The FIG design can give overly diffuse marginal posteriors for some parameters.
The ADV design avoids this drawback, and appears to be similar to the SIG design,
and indeed gives competitive performance under the SIG objective.
Multiple runs of ADV produced designs with similar cluster locations but varying cluster sizes.
Post-processing these designs using point exchange further improved the ADV objective reached,  illustrating the importance of this step.
To explain the improvement, note that point exchange found two candidates for optimal cluster sizes, including one which was different from any cluster sizes found without post-processing.

\subsection{Realistic example} \label{sec:pk_realistic}

Here we implement our ADV approach on a more realistic version of the pharmacokinetic example.
Firstly, we now assume \emph{multiplicative noise}:
\begin{equation} \label{eq:multi}
y_i \sim N(x(\theta, \tau_i), \sigma_1^2 + \sigma_2^2 x(\theta, \tau_i)^2),
\end{equation}
with $\sigma_1^2 = 0.1, \sigma_2^2 = 0.01$.
Secondly we require \emph{gaps} i.e.~a constraint that observation times are at least $0.25$ hours apart.
These changes result in the model used by \cite{Ryan:2014} and \cite{Overstall:2017}.

Results exist for the FIM of a model with multiplicative noise
(e.g.~see \citealp{Malago:2015}, equation 23).
However a lengthy analytic derivation is required which can easily result in errors.
Therefore we implement a method which can be used when the FIM is intractable:
Algorithm \ref{alg:GDA_BED2}, described in Appendix \ref{sec:estimationMore}.
Appendix \ref{sec:realistic_supp} gives further implementation details for this example and more comments on the results.

\begin{figure}[htp]
\begin{center}
\includegraphics[width=0.7\textwidth]{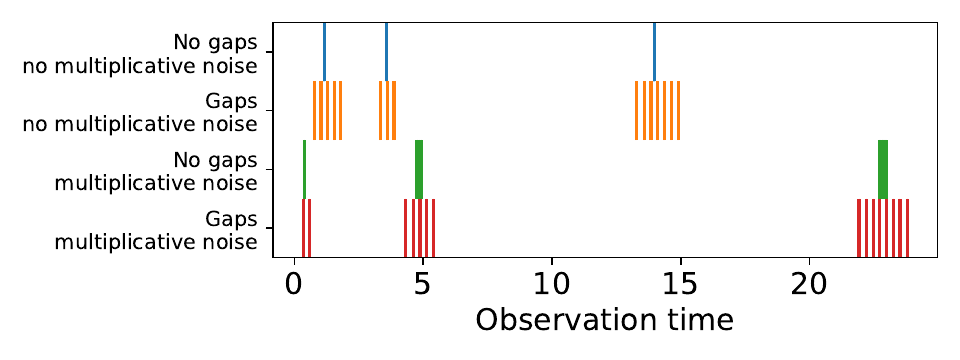}
\end{center}
\caption{Optimal ADV designs for several variations of the pharmacokinetic example.}
\label{fig:pk_designs}
\end{figure}

Figure \ref{fig:pk_designs} shows the resulting designs.
Without the gaps constraint, there are 3 clusters of repeated design points.
The cluster locations differ depending on whether or not multiplicative noise is used.
When gaps are enforced, the clusters remain at the same locations, but consist of spaced out, rather than repeated, design points.

\section{Geostatistical regression example} \label{sec:geostat}

This section considers an example requiring hundreds of design choices,
to illustrate how our method can scale up to higher dimensional applications.

\subsection{Model}

Consider the following geostatistical regression model.
Here a design $\tau$ is a $d \times 2$ matrix whose rows specify measurement locations.
We assume normal observations with a linear trend and squared exponential covariance function with a nugget effect, giving
\begin{align*}
y &\sim N(x(\theta, \tau), \Sigma(\tau)), \quad
&x_i &= \theta_1 \tau_{i1} + \theta_2 \tau_{i2}, \\
\Sigma &= \sigma_1^2 I + \sigma_2^2 R(\tau), \quad
&R_{ij} &= \exp \left[-\sum_{k=1}^2 (\tau_{ik} - \tau_{jk})^2 / \ell^2 \right].
\end{align*}
For simplicity we assume that $\sigma^2_1, \sigma^2_2$ (observation variance components) and $\ell$ (covariance length scale) are known.
Hence the unknown parameters are $\theta_1$ and $\theta_2$ (trends).

\subsection{Methods}

Using \eqref{eq:mvn_info} the FIM is
$\mathcal{I}(\tau) = \tau^T \Sigma(\tau)^{-1} \tau$,
which does not depend on $\theta$.
Hence $\mathcal{I}(\tau)$ is also the expected FIM
and we do not need to use Monte Carlo to estimate it.

We performed simulation studies
for $\ell = 0.01,0.02,0.04,0.08$ with $\sigma_1=1$ and $\sigma_2=3$
to search for $500$ design points restricted to a unit square centred at the origin.
The design was initialised as independent uniform draws.
Each run used Algorithm \ref{alg:GDA_BED} with the Adam update rule for 1000 iterations,
which was enough for convergence (see Appendix \ref{app:geostats}).
We implemented constrained optimisation by adding a $L_1$ penalty to designs outside the unit square.

As this example aims to illustrate the time required by ADV, we did not investigate repeated runs from different initial designs, or post-processing using point exchange.
In any case, the latter seems unlikely to help as there is little evidence of replicated observations in the results.

\subsection{Results}

\begin{figure}[htbp]
  \centering
  \setlength\tabcolsep{0pt} 
  \begin{tabular}{cccc}
    $\ell=0.01$ & $\ell = 0.02$ & $\ell = 0.04$ & $\ell = 0.08$ \\
    \begin{minipage}{0.25\textwidth} \includegraphics[width =\textwidth]{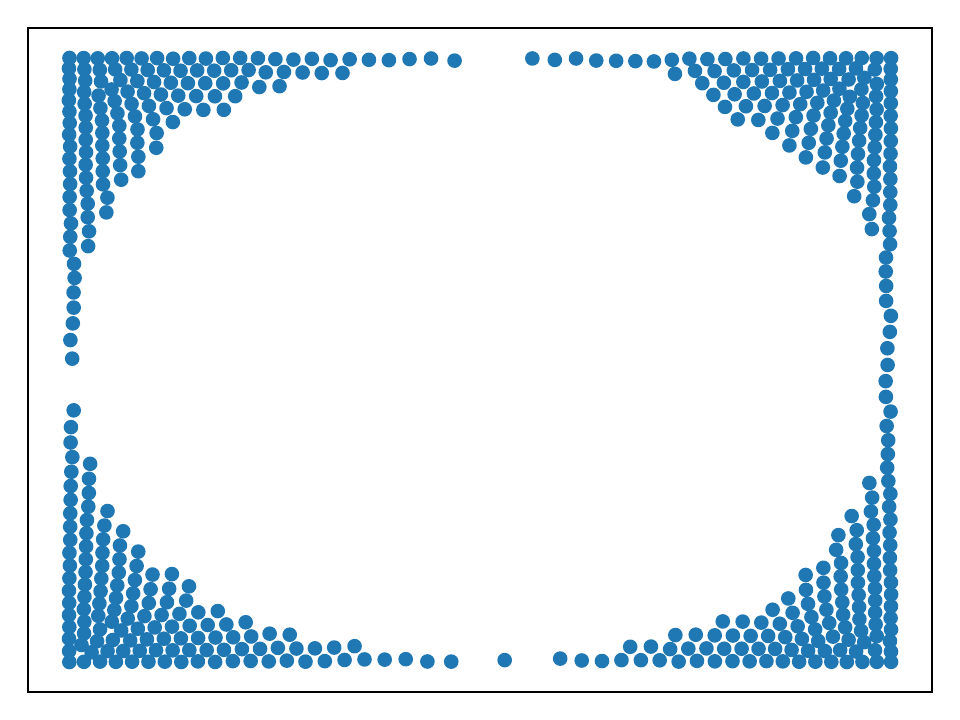} \end{minipage} &
    \begin{minipage}{0.25\textwidth} \includegraphics[width =\textwidth]{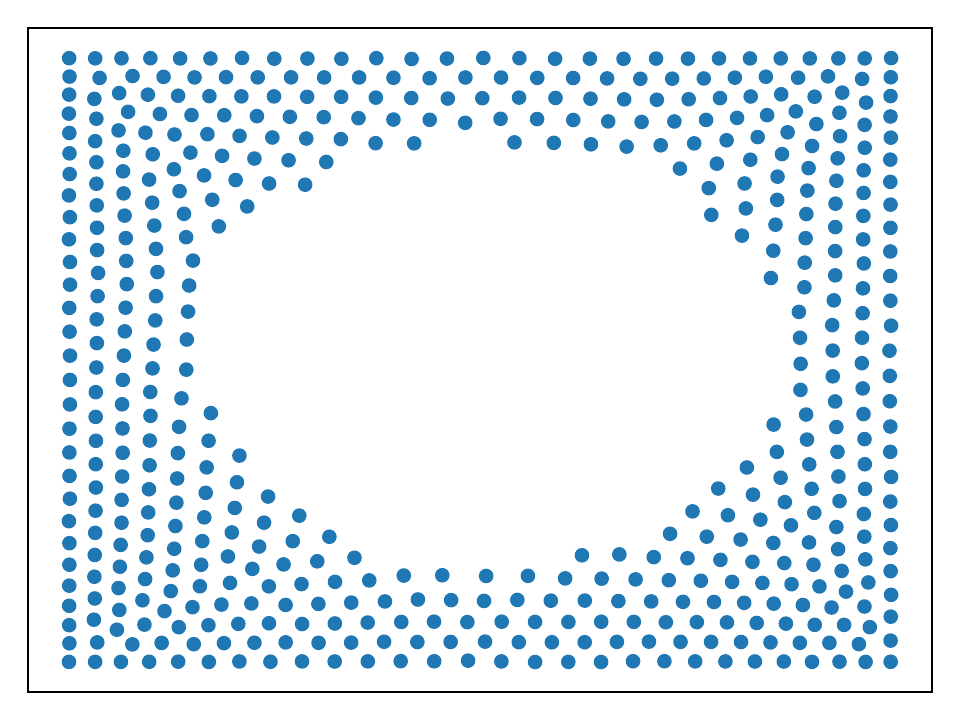} \end{minipage} &
    \begin{minipage}{0.25\textwidth} \includegraphics[width =\textwidth]{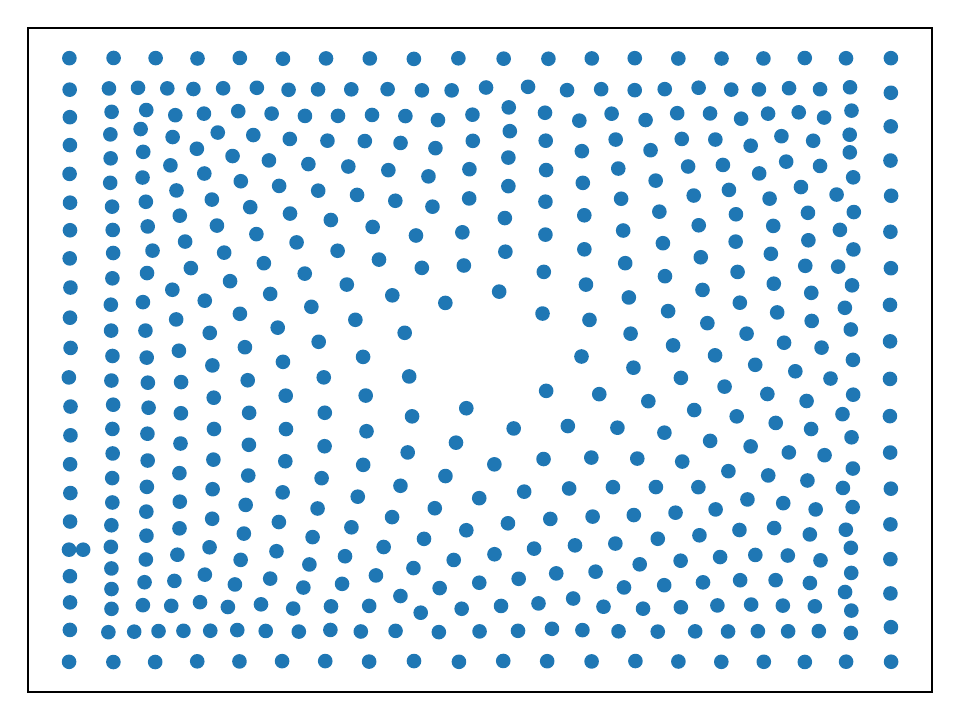} \end{minipage} &
    \begin{minipage}{0.25\textwidth} \includegraphics[width =\textwidth]{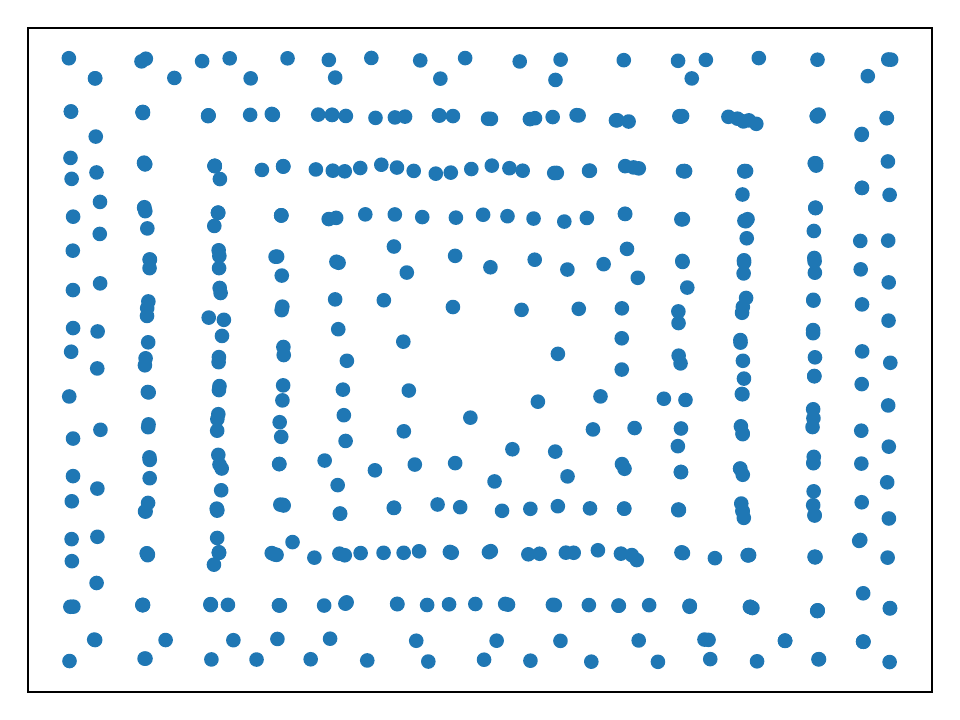} \end{minipage}
  \end{tabular}
  \caption{Geostatistical regression model designs returned for various choices of $\ell$.
The design space is a unit square centred at the origin.}
  \label{fig:geodesigns}
\end{figure}

Optimisation took on average 19.4 seconds.
Figure \ref{fig:geodesigns} shows the resulting designs.
For small $\ell$ values, the design points cluster in the corners.
For larger values, the designs are spread across the region
with varying spatial structures.
For all runs $A$ remained very close to the identity matrix throughout optimisation,
reflecting the symmetry of $\theta_1$ and $\theta_2$ in the model.
Hence FIG would also work well for this example.

\section{Discussion} \label{sec:discussion}

We have presented a gradient descent ascent algorithm Bayesian optimal design using an objective function based on the Fisher information.
This provides improvements in speed and scalability to higher dimensional designs by avoiding the need for posterior inference.
We also provide a novel game theoretic justification for our objective,
and provide theoretical insights into the choice of utilities for Bayesian optimal design from decision/game theoretic principles,
extending the work of \cite{Bernardo:1979} and \cite{Walker:2016}.

In simulation studies our approach finds locally optimal design faster than other state of the art methods, by a factor of at least 10.
To assess whether multiple locally optimal designs exist we recommend starting from multiple random initial designs, which can easily be done in a parallel version of our algorithm.
If locally optimal designs involve clusters of repeated observations,
we recommend post-processing using a point exchange algorithm.
Although we did not observe any in our examples,
GDA dynamics can converge to limit cycles rather than the desired solutions.
We recommend checking for cyclic behaviour using trace plots of $\tau$ values produced during optimisation.

\subsection{Limitations and future research}

\paragraph{Intractable Fisher information}

Our methods have assumed that the FIM, and associated gradients, can easily be evaluated.
When the FIM cannot be evaluated but the likelihood or score function \eqref{eq:score} can, it is possible to produce an unbiased estimate of the FIM.
Appendix \ref{sec:estimationMore} describes this and how it can be used for experimental design.
Section \ref{sec:pk_realistic} contained an application.

However sometimes the likelihood and score function cannot be evaluated.
One common reason is the presence of latent variables, such as nuisance parameters.
Appendix \ref{sec:estimationMore} describes difficulties of implementing
our approach in this setting,
and sketches methods to do so, which we plan to investigate in future research.

\paragraph{Discrete designs}

Gradient based optimisation is only available for a continuous space of possible designs.
It would be of interest to develop analogous methods to this paper for discrete designs.
These could be based on discrete optimisation algorithms,
or involve relaxation of the discrete problem to a continuous approximation.

\paragraph{Discrete parameters}

A limitation of our work is that it does not apply to discrete parameters,
since Hyv\"arinen score and FIM are only defined for continuous parameters.
Discrete analogues have been proposed \citep{Dawid:2012, Shao:2019},
which would be interesting to investigate.



\paragraph{Variance reduction}

Optimisation efficiency could be increased by reducing the variance of our Monte Carlo gradient estimates.
For instance, a reviewer suggests the use of randomised quasi-Monte Carlo, as in \cite{Drovandi:2018}.

\paragraph{Alternative optimisers}

Variations on GDA with better convergence guarantees,
such as two time scaled update rules \citep{Heusel:2017},
are an active topic of research and could be used for the objective in this paper.
Another possibility for future work is modifying generic gradient based optimisation methods to avoid local optima in optimal design problems
e.g.~using tempering methods, or non-local updates such as line search, as used by \cite{Overstall:2017}.

\paragraph{Variations to game theoretic framework}

Several details of our game theoretic framework could be altered.
One possibility is to consider alternative scoring rules.
Many alternative proper scoring rules exist beyond logarithmic and Hyv\"arinen \citep{Parry:2012}.
Alternatively, non-proper scoring rules could be used which emphasise a particularly important aspect of inference to the task at hand e.g.~the tails of the distribution \citep{Loaiza:2019}.
Another variation is to allow more freedom to the critic.
This could include the ability to make more general, non-linear, reparameterisations, or to condition their reparameterisation on the observations, $y$.

\section*{Acknowledgements} 

The authors thank Kevin Wilson, Chris Oates, John Matthews, Phil Dawid, Kyle Cranmer, Iain Murray and journal referees/editors for helpful comments, and gratefully acknowledge an EPSRC studentship supporting Sophie Harbisher.

\bibliographystyle{apalike}
\bibliography{advFIG}

\clearpage

\appendix

\section{Scoring rule details} \label{app:scoring}

Section \ref{sec:scoring rules} of the main paper describes scoring rules and related concepts.
Here, Table \ref{tab:scoring} summarises two strictly proper scoring rules of particular interest in this paper:
logarithmic score and Hyv\"arinen score \citep{Hyvarinen:2005}.
It also gives their associated entropies and divergences.
The Hyv\"arinen results rely on the following regularity conditions:
\begin{enumerate}  
\item[H1.]
$p(\theta)$ and $q(\theta)$ are twice differentiable with respect to all $\theta_i$.
\item[H2.]
$\E_{\theta \sim p(\theta)}[\, || \nabla \log p(\theta) ||^2 ], \E_{\theta \sim q(\theta)}[\, || \nabla \log q(\theta) ||^2 ]$ are finite.
\item[H3.]
$\nabla p(\theta) \to 0, \nabla q(\theta) \to 0$ for $||\theta|| \to \infty$.
\end{enumerate}
It this section we use $\nabla$ without a subscript to represents gradient with respect to $\theta$.
Also recall that $||\theta||$ represents the $L_2$ norm.
Finally, note that several of the results of this paper use the following assumption.
\begin{enumerate}
\item[A5]
Conditions H1-H3 hold for $p(\theta) = \pi(\theta | y; \tau)$ and $q(\theta) = \pi(\theta)$, given any $y,\tau$.
\end{enumerate}

\begin{table}[hbtp]
\centering
\begin{tabular}{@{} lll@{}}
  \toprule
                                                                      & Logarithmic score                             & Hyv\"arinen score                                                        \\
  \midrule
Scoring rule                                                          & $-\log q(\theta)$                             & $2 \Delta \log q(\theta) + ||\nabla \log q(\theta)||^2$                  \\
Entropy                                                               & $-\E_{\theta \sim p(\theta)}[\log p(\theta)]$ & $-\E_{\theta \sim p(\theta)} \left[\,|| \nabla \log p(\theta) ||^2\right]$ \\
Divergence                                                            & $\E_{\theta \sim
                                                          p(\theta)}[\log
                                                          p(\theta) - \log
                                                          q(\theta)]$ & $\E_{\theta \sim p(\theta)} [\,||\nabla \log p(\theta) - \nabla \log q(\theta)||^2] $                                     \\
  \bottomrule
\end{tabular}
\caption{Summary of two scoring rules and related quantities.
Here $p(\theta)$ is the true density of an unobserved quantity and $q(\theta)$ is a generic density.
Note that $\Delta$ is the Laplacian operator (sum of second partial derivatives) with respect to $\theta$.
For a derivation of the Hyv\"arinen divergence see Appendix A of \cite{Hyvarinen:2005}.
The other derivations are straightforward.}
\label{tab:scoring}
\end{table}

\section{Framework summaries} \label{app:frameworks}

Table \ref{tab:dec} summarises and contrasts the decision-theoretic and game-theoretic frameworks from Sections \ref{sec:DTtheory} and \ref{sec:GTtheory}.

\begin{table}[bthp]
\begin{tabular}{p{0mm}p{60mm}|p{0mm}p{70mm}}
& Decision theoretic && Game theoretic \\
\toprule[1pt]
\textbullet & \textbf{E} selects design $\tau$ &
\textbullet & \textbf{E} selects design $\tau$ \\
&& \textbullet & \textcolor{blue}{\textbf{C} selects matrix $A$} \\
\textbullet & \textbf{N} generates parameters $\theta$ from prior &
\textbullet & \textbf{N} generates parameters $\theta$ from prior \\
& (unobserved until reward allocated) && (unobserved until reward allocated) \\
\textbullet & \textbf{N} generates data $y$ from model &
\textbullet & \textbf{N} generates data $y$ from model \\
\textbullet & \textbf{E} selects $a$, estimated density for $\theta$ &
\textbullet & \textcolor{blue}{\textbf{E} selects $a$, estimated density for $\phi = A^{-1} \theta$} \\
\textbullet & \textbf{E} receives reward $\mathcal{R}(a,\theta)$ &
\textbullet & \textcolor{blue}{\textbf{E} receives reward $\mathcal{R}(a,\phi)$} \\
&&
\textbullet & \textcolor{blue}{\textbf{C} receives reward $-\mathcal{R}(a,\phi)$} \\
\bottomrule[1pt]
\end{tabular}
\caption{Two frameworks for experimental design:
the approach of \cite{Bernardo:1979} (left),
and our modification (right), with changes shown in blue.
The abbreviations are for \textbf{E}xperimenter, \textbf{C}ritic and \textbf{N}ature.
Players observe all actions except $\theta$.
}
\label{tab:dec}
\end{table}

\section{Proofs} \label{sec:proofs}

\subsection{Proof of Result \ref{res:DTmain}} \label{sec:proofDTmain}

We will show the following which suffices to give the required result.
As in the statement of Result \ref{res:DTmain}, all expectations are with respect to $\pi(\theta, y; \tau)$.
\begin{enumerate}
\item
The expected reward from design $\tau$ equals $\E[\mathcal{U}_{\text{entropy}}]$,
assuming the experimenter selects $a$ to maximise their expected utility.
\item
$\E[\mathcal{U}_{\text{entropy diff}}]$ and $\E[\mathcal{U}_{\text{divergence}}]$ also equal this value up to an additive constant.
\end{enumerate}

Proofs of both parts are below.
The first is a consequence of A1.
The proof of the second part is based on the common observation that the divergence contains a constant term which can be ignored under maximisation.
Similar results have appeared in the experimental design literature previously for a logarithmic score function \cite[e.g.][]{Shewry:1987}.
This result simply generalises them to a general score function.

Note that manipulations of expectations used below are valid by A2 and Fubini's theorem.

\paragraph{Part 1}

First fix some values of $\tau$ and $y$ and consider the experimenter's choice of $a$.
This must maximise the expected reward, which is, using A1,
\[
-\E_{\theta \sim \pi(\theta | y; \tau)}[\mathcal{S}(a,\theta)].
\]
Since $\mathcal{S}$ is a strictly proper scoring rule,
the optimal choice of $a$ is the posterior $\pi(\theta | y; \tau)$.
The resulting expected reward is $\mathcal{U}_{\text{entropy}}(\tau, y)$.
(We write $\mathcal{U}_{\text{entropy}}$ having only the arguments $\tau, y$ as it does not depend on $\theta$.)

Now suppose $y$ is randomly generated by nature.
Then the expected reward of design $\tau$ is the expectation of $\mathcal{U}_{\text{entropy}}(\tau, y)$ with respect to $f(y|\theta,\tau)$,
or equivalently with respect to $\pi(\theta, y; \tau) = \pi(\theta) f(y|\theta;\tau)$, as required.

\paragraph{Part 2}

From our definitions,
\[
\mathcal{U}_{\text{divergence}} = \mathcal{U}_{\text{entropy}} + \E_{\theta \sim \pi(\theta | y; \tau)}[S(\pi(\theta), \theta)].
\]
We are interested in expectations of utility with respect to $\pi(\theta, y; \tau)$,
which equals $\pi(\theta | y; \tau) \pi(y; \tau)$ by \eqref{eq:joint}. Thus
\begin{align*}
\E_{(\theta,y) \sim \pi(\theta, y; \tau)} [\mathcal{U}_{\text{divergence}}]
&= \E_{(\theta,y) \sim \pi(\theta, y; \tau)} [\mathcal{U}_{\text{entropy}}]
+ \E_{(\theta,y) \sim \pi(\theta, y; \tau)}[S(\pi(\theta), \theta)] \\
&= \E_{(\theta,y) \sim \pi(\theta, y; \tau)} [\mathcal{U}_{\text{entropy}}]
+ \mathcal{H}[\pi(\theta)] \\
&= \E_{(\theta,y) \sim \pi(\theta, y; \tau)} [\mathcal{U}_{\text{entropy diff}}].
\end{align*}
(Note A2 guarantees the terms above are finite.)

So $\mathcal{U}_{\text{divergence}}$ and $\mathcal{U}_{\text{entropy diff}}$ produce the same expected utility.
Furthermore the expected utility of $\mathcal{U}_{\text{entropy}}$ only differs by a finite additive constant, $\mathcal{H}[\pi(\theta)]$.

\subsection{Proof of Result \ref{res:SIG/FIG}} \label{sec:proofSIGFIG}

It's sufficient to show that these utilities have the same expectation as $\mathcal{U}_{\text{divergence}}$ with respect to $\pi(\theta, y; \tau)$.
From Table \ref{tab:scoring} (and using A5 for the Hyv\"arinen score case):
\[
\mathcal{U}_{\text{divergence}} = 
\begin{cases}
\E_{\theta \sim \pi(\theta|y;\tau)}[\log \pi(\theta|y;\tau) - \log \pi(\theta)]
& \text{(log score)} \\
\E_{\theta \sim \pi(\theta|y;\tau)}[\,||\nabla_\theta \log \pi(\theta|y;\tau) - \nabla_\theta \log \pi(\theta)||^2\,]
& \text{(Hyv\"arinen score)}
\end{cases}
\]
The quantities within the expectations are $\mathcal{U}_{\text{SIG}}$ and $\mathcal{U}_{\text{FIG}}$ respectively.
Thus these have same expectations as $\mathcal{U}_{\text{divergence}}$ under $\pi(\theta, y; \tau)$.
Finally note that from \eqref{eq:FisherInfo},
$\mathcal{U}_{\text{trace}} = \E_{y \sim f(y|\theta;\tau)} [\mathcal{U}_{\text{FIG}}]$.
So this has the same expectation under $\pi(\theta, y; \tau)$ as $\mathcal{U}_{\text{FIG}}$.
(These manipulations of expectations are valid using Fubini's theorem and assumption A2 for the log score case, or A5 for the Hyv\"arinen score case.)

\subsection{Proof of Result \ref{res:GTlogscore}} \label{sec:proofGTlogscore}

As the logarithmic score is a proper scoring rule, $a(\tau,A,y)$ must output the posterior.
Using the change of variables formula,
the posterior for $\phi = A^{-1} \theta$ is
\begin{equation} \label{eq:GTlogscoreproof}
\pi_\Phi(\phi | y; \tau) = \pi_\Theta(\theta | y; \tau) |\det A|,
\end{equation}
where $\pi_\Theta(\theta | y; \tau)$ is the posterior for $\theta$.
Using A3, $\det A = 1$ so $\pi_\Phi(\phi | y; \tau) = \pi_\Theta(\theta | y; \tau)$.
Hence the expected reward to the experimenter from $\tau$ given $a(\tau,A,y)$ is,
regardless of the choice of $A$,
$\E_{\theta,y \sim \pi(\theta, y; \tau)}[-\log \pi_\Theta(\theta | y; \tau)]$,
the same as in the decision theoretic framework.
Thus $\tau$ must optimise the same objective in the decision theoretic setting,
or in a SPE of the game theoretic setting.

\subsection{Proof of Result \ref{res:GTreparam}} \label{sec:proofGTreparam}

Let ``game 1'' be based on the original parameters $\theta$,
and ``game 2'' use the alternative parameters $B \theta$.

\subsubsection{Part 1: Relation between action sequences}

Here we show that there are invertible mappings $A_2=f(A_1), a_2=g(a_1)$ such that actions $(\tau,A_1,y,a_1)$ in game 1, and $(\tau,A_2,y,a_2)$ in game 2 have the same expected rewards up to a multiplicative constant that depends only on $B$.
In game 1, $a_1$ is a density $a_1(\phi_1)$ for $\phi_1 = A_1^{-1} \theta$.
The expected reward to the experimenter is then $r_1 = -\E_{\theta \sim \pi(\theta | y; \tau)}[\mathcal{S}(a_1, A_1^{-1} \theta)]$
where $\mathcal{S}$ is Hyv\"arinen score.

First suppose $\det B > 0$.
In this case let $A_2 = k^{-1/p} B A_1$ where $k = \det B$ and $p = \dim(\theta)$.
This is a valid action in game 2 since $\det A_2 = 1$.
In game 2 the experimenter must select a density on
$\phi_2 = A_2^{-1} B \theta = k^{1/p} \phi_1$.
Let $a_2(\phi_2) = a_1(k^{-1/p} \phi_2) / k$.
Now consider the actions $(\tau, A_2, y, a_2)$ in game 2.
The expected reward to the experimenter is then $r_2 = -\E_{\theta \sim \pi(\theta | y; \tau)}[\mathcal{S}(a_2, k^{1/p} A_1^{-1} \theta)]$.
From the definition of Hyv\"arinen score, $r_2 = k^{-2/p} r_1$.
(As reparameterising by a scalar preserves Hyv\"arinen score up to proportionality.)

Now suppose $\det B < 0$. (Recall $B$ is invertible so $\det B \neq 0$.)
Let $F$ be a $p \times p$ identity matrix with the top left entry changed to $-1$.
Now let $A_2 = k^{-1/p} B A_1 F$ where $k = |\det B|$.
Again $\det A_2 = 1$ so this is a valid action in game 2.
In game 2 the experimenter must select a density on
$\phi_2 = A_2^{-1} B \theta = k^{1/p} F \phi_1$.
Let $a_2(\phi_2) = a_1(k^{-1/p} F \phi_2) / k$.
Now consider the actions $(\tau, A_2, y, a_2)$ in game 2.
The expected reward to the experimenter is then $r_2' = -\E_{\theta \sim \pi(\theta | y; \tau)}[\mathcal{S}(a_2, k^{1/p} F A_1^{-1} \theta)]$.
From the definition of Hyv\"arinen score, $r_2' = k^{-2/p} r_1$.
(As Hyv\"arinen score is preserved under reparameterisation by axis-aligned reflections.)

\subsubsection{Part 2: Relation between subgame perfect equilibria}

Consider a SPE of game 1 with decision rules $\tau$, $A(\tau)$, $a(\tau,A,y)$.
Define $A'(\tau) = f(A(\tau))$ and $a'(\tau,A,y) = g(a(\tau,A,y))$.
Then $\tau$, $A'(\tau)$, $a'(\tau,A,y)$ is a SPE of game 2.
This follows since any counterexample could be transformed to a counterexample for the original SPE in game 1.
Similarly a SPE of game 2 can be transformed to a SPE of game 1 using the inverse mappings.
This suffices to prove the main result.

\subsection{Proof of Result \ref{res:GTmain}} \label{sec:proofGTmain}

\subsubsection{Part 1: Objective function for $\tau, A$}

Initially the proof follows that of Results \ref{res:DTmain} and \ref{res:SIG/FIG}.
First consider the choice of $a(\tau,A,y)$ in a SPE.
This must maximise the expected reward to the experimenter, which is, using A1,
$-\E_{\phi \sim \pi_\Phi(\phi | y; \tau)}[\mathcal{S}(a,\phi)]$.
Since $\mathcal{S}$ is a strictly proper scoring rule,
the optimal choice of $a$ is the posterior $\pi_\Phi(\phi | y; \tau)$.
Hence the optimal $a(\tau,A,y)$ outputs
$\pi_\Phi(\phi | y; \tau)$,
and the resulting expected reward to the experimenter is the negative posterior entropy
\[
\E_{\phi \sim \pi_\Phi(\phi | y; \tau)}[\mathcal{R}(\pi_\Phi(\phi | y; \tau), \phi)] = -\mathcal{H}[ \pi_\Phi(\phi | y; \tau) ].
\]
The expected reward to the experimenter for this $a$ decision rule under random $y$ is
\begin{equation}
\E_{\phi, y \sim \pi_\Phi(\phi, y; \tau)}[ -\mathcal{H}[ \pi_\Phi(\phi | y; \tau) ] ]
= \E_{\phi, y \sim \pi_\Phi(\phi, y; \tau)}[ \mathcal{D}[\pi_\Phi(\phi|y;\tau), \pi_\Phi(\phi)] - \mathcal{H}[ \pi_\Phi(\phi) ] ]. \label{eq:rewardSum}
\end{equation}
(Noting that A5 implies that the terms above are finite and allows manipulation of expectations using Fubini's theorem.)
Define $\mathcal{K}(\tau,A)$ to be the negative of the first term on the right hand side of \eqref{eq:rewardSum}.
Then
\begin{align*}
\mathcal{K}(\tau,A) &= -\E_{\phi, y \sim \pi_\Phi(\phi, y; \tau)}[ \mathcal{D}[\pi_\Phi(\phi|y;\tau), \pi_\Phi(\phi)] ] \\
&= -\E_{\phi, y \sim \pi_\Phi(\phi, y; \tau)}[\,||\nabla_\theta \log \pi_\Phi(\phi|y;\tau) - \nabla_\theta \log \pi_\Phi(\phi)||^2\,] && \text{using Table \ref{tab:scoring}} \\
&= -\E_{\phi \sim \pi_\Phi(\phi), y \sim f_\Phi(y | \phi; \tau)}[\,||\nabla_\theta \log f(y|\phi;\tau)||^2\,] && \text{using \eqref{eq:posterior}} \\
&= -\E_{\phi \sim \pi_\Phi(\phi)} \trace [ \mathcal{I}_\phi(\phi; \tau) ] && \text{using \eqref{eq:FisherInfo}}.
\end{align*}
Note that the second line requires assumption A5.
Since $\phi = A^{-1} \theta$, applying \eqref{eq:reparam} gives
\begin{equation} \label{eq:Aobjective2}
\mathcal{K}(\tau,A)
= -\E_{\theta \sim \pi(\theta)} \trace [ A^T \mathcal{I}(\theta; \tau) A ]
= -\trace [ A^T \bar{\mathcal{I}}(\tau) A ].
\end{equation}
Here and in the remainder of the proof, $\pi$ and $\mathcal{I}$ without subscripts represent densities and FIM with respect to $\theta$.

Now consider the second term in \eqref{eq:rewardSum}, $\mathcal{H}[ \pi_\Phi(\phi) ]$.
From Table \ref{tab:scoring} and assumption A5
\[
\mathcal{H}[ \pi_\Phi(\phi) ] = -\E_{\theta \sim \pi(\theta)} \left[\,|| \nabla_\phi \log \pi_\Phi(\phi) ||^2\right],
\]
where $\phi = A^{-1} \theta$.
Using the change of variables formula
\begin{align*}
\log \pi_\Phi(\phi) &= \log \pi_\Theta(A \phi) + \log |\det A | \\
\Rightarrow \nabla_\phi \log \pi_\Phi(\phi) &= \nabla_\theta \log \pi_\Theta(\theta) | \det A | \\
&= \nabla_\theta \log \pi_\Theta(\theta) & \text{since } \det A = 1.
\end{align*}
Thus $\mathcal{H}[ \pi_\Phi(\phi) ]$ equals $\mathcal{H}[ \pi_\Theta(\theta) ]$ regardless of $A$.

So under the optimal decision rule $a(\tau,A,y)$,
the expected reward to the experimenter is $-\mathcal{K}(\tau,A)$, up to an additive constant.
Thus the remainder of the nested optimisation problem to be solved to identify SPEs is, as claimed,
\begin{equation} \label{eq:minmax}
\min_\tau \max_A \mathcal{K}(\tau,A).
\end{equation}

\subsubsection{Part 2: Lower bound on $\det \bar{\mathcal{I}}(\tau^*)$}

Here we prove that in a SPE, $\det \bar{\mathcal{I}}(\tau) > 0$.
This will be needed in part 4.
To show this result, we consider $\tau^*, A^*$ satisfying \eqref{eq:minmax} such that $\det \bar{\mathcal{I}}(\tau^*) = 0$, and derive a contradiction.

Firstly, consider the case $\mathcal{K}(\tau^*, A^*) = 0$.
By A4 there exists some $\tilde{\tau}$ such that $\det \bar{\mathcal{I}}(\tilde{\tau}) > 0$.
Then
\[
\det[ {A^*}^T \bar{\mathcal{I}}(\tilde{\tau}) A^* ]  = \det \bar{\mathcal{I}}(\tilde{\tau}) > 0.
\]
So all the eigenvalues of ${A^*}^T \bar{\mathcal{I}}(\tilde{\tau}) A^*$ are positive.
Hence
\[
\mathcal{K}(\tilde{\tau}, A^*) = -\trace[ {A^*}^T \bar{\mathcal{I}}(\tilde{\tau}) A^* ]
< 0
= \mathcal{K}(\tau^*, A^*),
\]
which contradicts $\tau^*, A^*$ satisfying \eqref{eq:minmax}.

Secondly, consider the case $\mathcal{K}(\tau^*, A^*) = r < 0$.
Express $\bar{\mathcal{I}}(\tau^*)$ as an eigendecomposition $P \Lambda P^T$.
Let $z$ be the number of zero eigenvalues (so there are $p-z$ non-zero eigenvalues).
We must have $z>0$ since $\det \bar{\mathcal{I}}(\tau^*) = 0$.
Let $\tilde{\Lambda}(\lambda)$ be matrix resulting from
replacing the zero eigenvalues in $\Lambda$ with $\lambda^{1/z}$
and the non-zero eigenvalues with $\lambda^{-1/(p-z)}$.
Define $\tilde{A}(\lambda) = P \tilde{\Lambda}(\lambda) P^T$.
Then $\det \tilde{A}(\lambda) = 1$ so it is a valid action.
Also $\lim_{\lambda \to \infty}\trace [ \tilde{A}(\lambda)^T \bar{\mathcal{I}}(\tau^*) \tilde{A}(\lambda) ] = 0$,
so for $\lambda$ sufficiently large
\[
\mathcal{K}(\tau^*, \tilde{A}(\lambda))
= -\trace[\tilde{A}(\lambda)^T \bar{\mathcal{I}}(\tau^*) \tilde{A}(\lambda)]
> r
= \mathcal{K}(\tau^*, A^*),
\]
which contradicts $\tau^*, A^*$ satisfying \eqref{eq:minmax}.

\subsubsection{Part 3: Optimal $\mathcal{K}$ given $\tau$}

Fix some $\tau$ such that $\det \bar{\mathcal{I}}(\tau) > 0$.
This part shows that $\max_A \mathcal{K}(\tau,A) = -p [\det \bar{\mathcal{I}}(\tau)]^{1/p}$.

From \eqref{eq:Aobjective2}, $\mathcal{K}(\tau,A) = -\text{tr} B$
where $B = A^T \bar{\mathcal{I}}(\tau) A$.
Let $\lambda_1,\ldots,\lambda_p$ be the eigenvalues of $B$.
These are all positive:
they are non-negative since $B$ is positive definite,
and non-zero since $\prod_{i=1}^p \lambda_i = \det \bar{\mathcal{I}}(\tau) > 0$.
Consider minimising $\trace B$
subject to $\det A = 1$, which implies $\det B = \det \bar{\mathcal{I}}(\tau)$.
Equivalently we must minimise $\sum_{i=1}^p \lambda_i$
subject to $\prod_{i=1}^p \lambda_i = \det \bar{\mathcal{I}}(\tau)$.
It is straightforward 
that the solution is $\lambda_i = [\det \bar{\mathcal{I}}(\tau)]^{1/p}$ for all $i$.
This can be realised by taking $A$ as a Cholesky factor of $[\det \bar{\mathcal{I}}(\tau)]^{1/p} \bar{\mathcal{I}}(\tau)^{-1}$.
The claimed result follows from this.

\subsubsection{Part 4: Conclusion}

We will partition the possible $\tau$ values into two types.
In both cases we will show that the set of those which appear in SPEs
equals the set of those which attain the maximum possible $\det \bar{\mathcal{I}}(\tau)$ value.

First consider $\tau$ such that $\det \bar{\mathcal{I}}(\tau) = 0$.
Such a $\tau$ does not appear in a SPE (by part 2)
and does not maximise $\det \bar{\mathcal{I}}(\tau)$ (from A4).
Hence the two sets are empty for this case.

Next consider the case of $\tau$ such that $\det \bar{\mathcal{I}}(\tau) > 0$.
Part 1 shows that SPE actions for $\tau, A$ are solutions to $\min_\tau \max_A \mathcal{K}(\tau,A)$.
Part 3 shows that, for this case, $\max_A \mathcal{K}(\tau,A) = -p [\det \bar{\mathcal{I}}(\tau)]^{1/p}$.
Thus for this case, the set of designs which appear in SPEs equals those maximising $\det \bar{\mathcal{I}}(\tau)$,
as required.

\subsection{Proof of Result \ref{res:unbiased}} \label{sec:proofUnbiased}

From \eqref{eq:K} we have
\[
\nabla_z \mathcal{K}(\tau, A) = -\nabla_z \E_{\theta \sim \pi(\theta)} \trace [A^T \mathcal{I}(\theta; \tau) A].
\]
The regularity conditions allow interchange of differentiation and expectation (see Appendix \ref{sec:regularity}) so
\[
\nabla_z \mathcal{K}(\tau, A) = -\E_{\theta \sim \pi(\theta)} \nabla_z \trace [A^T \mathcal{I}(\theta; \tau) A].
\]
An unbiased Monte Carlo estimate of this is
\[
-\frac{1}{K} \sum_{k=1}^K \nabla_z \trace [A^T \mathcal{I}(\theta^{(k)}; \tau) A
= -\nabla_z \trace \left[A^T \left\{ \frac{1}{K} \sum_{k=1}^K \mathcal{I}(\theta^{(k)}; \tau) \right\} A \right],
\]
as required.

\section{Restricting $\det A$} \label{app:det}

This appendix justifies introducing assumption A3 in the game theoretic setting, which enforces $\det A = 1$.
First recall that we required $A$ to be invertible.
For non-invertible $A$, $\phi = A^{-1} \theta$ would not be defined, and this is essential in the description of the game theoretic framework.
Hence we must have $\det A \neq 0$.

Some further restriction on $A$ is necessary for a SPE to exist,
at least for the scoring rules we consider.
First consider the logarithmic score case.
From \eqref{eq:GTlogscoreproof} we have
\begin{equation} \label{eq:logscorereparam}
\log \pi_\Phi(\phi | y; \tau) = \log \pi_\Theta(\theta | y; \tau) + \log |\det A|,
\end{equation}
Without a further restriction on $A$, the critic can always improve their expected reward, $-\E_{\phi \sim \pi_\Phi(\phi | y; \tau)}[\log \pi_\Phi(\phi | y; \tau)]$, by replacing $A$ with $\lambda A$ for $\lambda<1$.
(This makes the posterior for $\phi$ flatter, benefiting the critic.)

Next consider the Hyv\"arinen score case,
and the critic's expected reward in a SPE after the experimenter plays $\tau$.
From Section \ref{sec:proofGTmain} (part 1) this is
\[
\mathcal{K}(\tau,A) = -\E_{\theta \sim \pi(\theta)} \trace [ A^T \mathcal{I}(\theta; \tau) A ].
\]
Again, without a further restriction on $A$, the critic can always improve their expected reward by replacing $A$ with $\lambda A$ for $\lambda<1$.
(This assumes that the expected reward is not already zero, which is a consequence of assumption A4.)

The above suggests that an additional restriction on $A$ is needed beyond $\det A \neq 0$.
Assumption A3, $\det A = 1$, is a natural restriction to use since $\det A$ appears in \eqref{eq:logscorereparam}.

\section{Point exchange algorithms} \label{sec:pointExchange}

As discussed in Section \ref{sec:main alg}
we often wish to find a design made up of multiple \emph{design points} from some region,
and the optimal design is a small number of clusters of repeated observations.
Gradient based optimisation can usually locate the optimal cluster locations well
(i.e.~determine where the clusters should be located),
and here we present a simple \emph{point exchange algorithm} \citep{Atkinson:2007, Overstall:2017} to find approximately optimal cluster sizes
(i.e.~determine how many design points should be at each location).

The method seeks to optimise one of the diagnostics introduced in Section \ref{sec:main alg},
\begin{align*}
\hat{\mathcal{J}}_{\text{FIG}}(\tau) &= \trace \left\{ \frac{1}{J} \sum_{j=1}^J \mathcal{I}(\tilde{\theta}^{(j)}; \tau) \right\}, \\
\text{or} \quad \hat{\mathcal{J}}_{\text{ADV}}(\tau) &= \det \left\{ \frac{1}{J} \sum_{j=1}^J \mathcal{I}(\tilde{\theta}^{(j)}; \tau) \right\}.
\end{align*}
These are Monte Carlo estimates of 
$\mathcal{J}_{\text{FIG}}(\tau)$ from \eqref{eq:Jfig} and
$\mathcal{J}_{\text{ADV}}(\tau)$ from \eqref{eq:Jadv}.
The $\tilde{\theta}^{(j)}$ terms are independent samples from the prior,
with the same samples being reused in all evaluations of
$\hat{\mathcal{J}}_{\text{FIG}}$ or $\hat{\mathcal{J}}_{\text{ADV}}$.
Our implementation uses $J=1000$.
We then seek to maximise the deterministic function
$\hat{\mathcal{J}}_{\text{FIG}}(\tau)$ or $\hat{\mathcal{J}}_{\text{ADV}}(\tau)$.

We perform several iterations of greedy optimisation.
The design $\tau^{(0)}$ is initialised as the output of gradient based optimisation.
Iteration $i$ considers every possible candidate design formed from $\tau^{(i-1)}$ by replacing one design point with another.
The candidate design optimising $\hat{\mathcal{J}}_{\text{FIG}}$ or $\hat{\mathcal{J}}_{\text{ADV}}$ is used as $\tau^{(i)}$.
The algorithm terminates if $\tau^{(i)} = \tau^{(i-1)}$ or if a maximum number of iterations is exceeded.

Note that this is a simple approximate optimisation method,
and there is scope for more sophisticated algorithms to be developed for use here.

\section{Regularity conditions} \label{sec:regularity}

Several of our derivations, such as Result \ref{res:unbiased},
require weak regularity conditions on $g(\theta, \tau)$ to allow interchange of differentiation and expectation i.e.
\begin{equation*}
\nabla_\tau \E_{\theta \sim \pi(\theta)}[g(\theta, \tau)]
= \E_{\theta \sim \pi(\theta)}[\nabla_\tau g(\theta, \tau)].
\end{equation*}
Sufficient regularity conditions are that for all $\tau$,
\begin{enumerate}
\item
$\E_{\theta \sim \pi(\theta)}[\, |g(\theta, \tau)| ]$ is finite
\item 
$g(\theta, \tau)$ is differentiable in each component of $\tau$
\item
$\E_{\theta \sim \pi(\theta)}[\, | \nabla_\theta g(\theta, \tau)| ]$ is finite
\end{enumerate}
where absolute value $|\cdot|$ acts elementwise.
The required equality then follows using Fubini's theorem.

\section{Intractable Fisher information} \label{sec:estimationMore}

This appendix considers how to apply our approach when the FIM is not available in a closed form that can easily be evaluated.

First we consider two cases where extending our approach is relatively straightforward, and give an algorithm which does so.
Section \ref{sec:reparam_trick} looks at the case where the score function is available,
and Section \ref{sec:intractable_score} at the case where the likelihood is available.
In the latter we also discuss scope for future work to improve the efficiency of the algorithm's implementation in code.

Secondly we discuss two more complex cases which we plan to consider in future work.
At the request of a reviewer, we sketch methods to do so, to show that this is feasible.
Section \ref{sec:latent} looks at the case where there are latent variables, such as nuisance parameters,
and Section \ref{sec:ilikelihood} at the case where the likelihood is intractable.

\paragraph{Recap}

Recall that our ADV approach is to find minimax solutions of
\begin{equation} \label{eq:K2}
\mathcal{K}(\tau,A) = -\E_{\theta \sim \pi(\theta)} \trace [ A^T \mathcal{I}(\theta; \tau) A ],
\end{equation}
using gradient descent ascent.
As in the main paper we take $A=A(\eta)$, following \eqref{eq:A(eta)}.
GDA requires unbiased estimates of $\nabla_z \mathcal{K}(\tau,A)$
for $z \in \{\tau, \eta\}$.

\subsection{Estimation using the score function} \label{sec:reparam_trick}

First we consider the case where the score function $u(y,\theta;\tau) = \nabla_\theta \log f(y|\theta; \tau)$ can be evaluated.
We show that it is possible to produce unbiased estimates of the required gradients, provided assumption A6, stated below, is met.
This allows us to perform optimisation using GDA,
as described below in Algorithm \ref{alg:GDA_BED2}.

\paragraph{Estimation of objective}

Recall that \eqref{eq:FisherInfo} defines
\[
\mathcal{I}_\theta(\theta; \tau) = \E_{y \sim f(y | \theta; \tau)} [ u(y, \theta; \tau) u(y, \theta; \tau)^T ].
\]
Substituting this into \eqref{eq:K2} gives
\begin{align}
\mathcal{K}(\tau,A) &= -\E_{\theta \sim \pi(\theta), y \sim f(y | \theta; \tau)} \trace [ A^T u(y, \theta; \tau) u(y, \theta; \tau)^T A ] \nonumber \\
&= -\E_{\theta \sim \pi(\theta), y \sim f(y | \theta; \tau)} [ u(y, \theta; \tau)^T A A^T u(y, \theta; \tau) ] \label{eq:K2.5} \\
&= -\E_{\theta \sim \pi(\theta), y \sim f(y | \theta; \tau)} [\, ||A^T u(y, \theta; \tau)||^2 ], \label{eq:K3}
\end{align}
which can easily be estimated unbiasedly by Monte Carlo.

\paragraph{Estimation of gradients}

Getting unbiased estimators of $\nabla_z \mathcal{K}(\tau,A)$ is more complicated.
This is because on taking the gradient of \eqref{eq:K3},
it is generally not possible to exchange the order of gradient and expectation operators, as the distribution for $y$ depends on $\tau$.
Therefore we use the \emph{pathwise derivative} approach
(see \citealp{Mohamed:2020} for a review)
based on the following assumption:
\begin{enumerate}
\item[A6]
The observations $y$ can be generated as a transformation
$y(\epsilon, \theta, \tau)$ of a random variable $\epsilon$
with fixed density $p(\epsilon)$.
\end{enumerate}
This gives
\[
\mathcal{K}(\tau,A) = -\E_{\theta \sim \pi(\theta), \epsilon \sim p(\epsilon)} [\, || A^T u(y(\epsilon, \theta, \tau), \theta; \tau) ||^2 ].
\]
Now it is possible to exchange expectation and differentiation, assuming appropriate regularity conditions (see Appendix \ref{sec:regularity}),
\begin{equation*}
\nabla_z \mathcal{K}(\tau,A) = -\E_{\theta \sim \pi(\theta), \epsilon \sim p(\epsilon)} [ \nabla_z || A^T u(y(\epsilon, \theta, \tau), \theta; \tau) ||^2 ].
\end{equation*}
Hence an unbiased Monte Carlo estimate of $\mathcal{K}(\tau,A)$ is
\begin{equation} \label{eq:Kest2}
\breve{\mathcal{K}}(\tau,A) =
-\frac{1}{K} \sum_{k=1}^K [ ||A^T u(y(\epsilon^{(k)}, \theta^{(k)}, \tau), \theta^{(k)}; \tau)||^2 ],
\end{equation}
where $(\theta^{(k)}, \epsilon^{(k)})$ are independent draws from $\pi(\theta) p(\epsilon)$.
Also, it follows from the above that
$\nabla_z \breve{\mathcal{K}}(\tau,A)$ is an unbiased estimate of $\nabla_z \mathcal{K}(\tau,A)$
for $z \in \{\tau, \eta\}$.

In some cases A6 does not hold, as $y$ cannot be represented as a suitable transformation of $\epsilon$.
This includes the case of discrete $y$ and other common distributions such as Beta and Gamma.
Various techniques are available to deal with these cases such as
taking a continuous approximation to discrete variables \citep{Maddison:2017}
or using implicit differentiation \citep{Figurnov:2018, Jankowiak:2018}.

\paragraph{Algorithm}

Algorithm \ref{alg:GDA_BED2} presents our approach when the score function can be evaluated, based on the use of pathwise derivative gradient estimates.

\begin{algorithm}[htbp] \caption{Gradient descent ascent for Bayesian experimental design using estimated Fisher information} \label{alg:GDA_BED2}
\begin{algorithmic}[1]
\STATE Input:
Number of samples to use in \eqref{eq:Kest2} $K$,
number of parallel replications to perform $R$,
initial values $\tau_1^i, \eta^i_1$ (for $i = 1,2,\ldots,R$),
update subroutines $h_\tau, h_\eta$.
\FOR{$t = 1,2,\ldots$}
\STATE Sample $\theta^{(k)}$ from the prior for $k = 1, 2, \ldots, K$.
\STATE Sample $\epsilon^{(k)}$ from $p(\epsilon)$ for $k = 1, 2, \ldots, K$.
\STATE Compute $g^i_{\tau,t}, g^i_{\eta,t}$, unbiased estimates of $-\nabla_\tau \mathcal{K}(\tau_t^i, A(\eta_t^i))$ and $\nabla_\eta \mathcal{K}(\tau_t^i, A(\eta_t^i))$ using automatic differentiation of \eqref{eq:Kest2} for all $i$.
\STATE Update estimates using $\tau^i_{t+1} = \tau^i_t + h_\tau(g^i_{\tau,t})$, $\eta^i_{t+1} = \eta^i_t + h_\eta(g^i_{\eta,t})$ for all $i$.
\ENDFOR
\end{algorithmic}
\end{algorithm}

\subsection{Estimation using the likelihood} \label{sec:intractable_score}

Here we consider the case where the likelihood can be evaluated easily but the score function cannot.
In this case whenever the score function is required it can simply be calculated by automatic differentiation of the log likelihood, 
and we can use Algorithm \ref{alg:GDA_BED2}.
Section \ref{sec:pk_realistic} has an example using this approach,
with further details in Appendix \ref{sec:realistic_supp}.

Our PyTorch code for this case is restricted to $R=K=1$
i.e.~number of parallel replications and Monte Carlo sample size must both equal one.
Otherwise an iteration of the algorithm requires calculating likelihood gradients for several $(\theta, y)$ combinations, then manipulating these, and finally calculating gradients again: it is difficult to perform such a calculation efficiently in PyTorch.
In future work it would be interesting to implement this method in other frameworks
which provide more flexibility in the use of automatic differentiation, such as Jax \citep{Bradbury:2018}.

\subsection{Latent variables case} \label{sec:latent}

Next we discuss a case where the likelihood cannot be evaluated as the observations depend on \emph{latent variables} $\psi$.
The latent variables could represent:
\begin{itemize}
\item
\emph{Nuisance parameters} that we do not wish to learn from our experiment, but which affect the distribution of $y$ nonetheless.
\item
\emph{Unobserved states} in a time series model.
\item
\emph{Model parameters} when we are interested only in model choice.
\end{itemize}
Below we sketch an approach to experimental design in the presence of latent variables.
This illustrates the difficulty of this case but that progress is feasible.
We plan to investigate this further in future work.

\paragraph{Estimation of score}

The observations now depend on latent variables $\psi$,
so their density is $f(y | \theta, \psi; \tau)$.
Let the conditional prior density of $\psi$ be $\pi(\psi | \theta)$.
We would like to work with the \emph{marginalised likelihood}
\[
f(y | \theta; \tau) = \int f(y | \theta, \psi; \tau) \pi(\psi | \theta) d\psi,
\]
and calculate the corresponding \emph{marginalised score function}
$\nabla_\tau \log f(y | \theta; \tau)$
and the resulting Fisher information matrix.
Unfortunately the marginalised likelihood involves an intractable integral,
which typically cannot be evaluated.

However an unbiased estimate of the marginalised score function exists,
using \emph{Fisher's identity} \citep{Cappe:2006, Poyiadjis:2011}
\begin{equation} \label{eq:Fisher_identity}
u(y,\theta; \tau) = \nabla_\theta \log f(y | \theta; \tau) = \E_{\psi \sim \pi(\psi | \theta, y; \tau)} [ \nabla_\theta \log f(y, \psi | \theta; \tau) ].
\end{equation}
where $\pi(\psi | \theta, y; \tau)$ is a posterior for the latent variables
conditional on $\theta$.

\paragraph{Difficulties}

The two difficulties in implementing the ADV approach with latent variables are:
\begin{enumerate}
\item Inference for $\pi(\psi | \theta, y; \tau)$.
\item Getting unbiased estimates of $\nabla_\tau u(y,\theta; \tau)$.
\end{enumerate}
The latter is required in estimating $\nabla_\tau \mathcal{K}(\tau,A)$,
and it is difficult because on taking the gradient of \eqref{eq:Fisher_identity},
it is not possible to exchange the order of gradient and expectation operators,
as the posterior distribution for $\psi$ depends on $\tau$.

Below we sketch an approach which deals with both difficulties.

\paragraph{Estimation of $\psi$ posterior}

For now, suppose $\tau$ is fixed.
We would like to have access to $\pi(\psi | \theta, y; \tau)$
for any choices of $\theta$ and $y$.
An approximate approach is to use a version of \emph{neural density estimation}.
The goal is to find a function $g(\varepsilon, \theta, y)$
which given $\varepsilon \sim N(0,I)$
outputs samples from a close estimate of $\pi(\psi | \theta, y; \tau)$.
The function $g$ can be a neural network of suitable form,
such as a \emph{normalising flow} \citep{Papamakarios:2021}.
The neural network's tuning parameters can be trained by using SGD
to maximise the likelihood based on samples of $(\theta, \psi, y)$ from the prior and model \citep{Papamakarios:2021}.
See \cite{Foster:2019, Foster:2020, Kleinegesse:2020} for approaches to Bayesian experimental design under the SIG objective
which make use of neural density estimation or closely related methods.

\paragraph{Estimation of objective gradients}

Suppose one has access to a function $g(\varepsilon, \theta, y)$
which given $\varepsilon \sim N(0,I)$ produces samples from $\pi(\psi | \theta, y; \tau^*)$ where $\tau^*$ is the optimal design.
Fisher's identity \eqref{eq:Fisher_identity} then becomes
\begin{equation} \label{eq:FI2}
u(y,\theta; \tau) = \E_{\varepsilon \sim N(0,I)} [ \nabla_\theta \log f(y, \psi | \theta; \tau) ].
\end{equation}
where $\psi$ is a shorthand for\footnote{
Note that $\epsilon$ and $\varepsilon$ represent different random vectors.
} $g(\varepsilon, \theta, y)$.
Similarly, we will assume A6 and write $y$ as shorthand for $y(\epsilon, \theta, \tau)$.
Substituting \eqref{eq:FI2} into \eqref{eq:K2.5} gives
\begin{align*}
\mathcal{K}(\tau,A) &=
-\E_{\theta \sim \pi(\theta), \epsilon \sim p(\epsilon),
\varepsilon_1 \sim N(0,I), \varepsilon_2 \sim N(0,I)}
[ \log f(y, \psi_1 | \theta; \tau)^T A A^T \log f(y, \psi_2 | \theta; \tau) ], \\
\nabla_z \mathcal{K}(\tau,A) &=
-\E_{\theta \sim \pi(\theta), \epsilon \sim p(\epsilon),
\varepsilon_1 \sim N(0,I), \varepsilon_2 \sim N(0,I)}
[ \nabla_z \left\{
\log f(y, \psi_1 | \theta; \tau)^T A A^T \log f(y, \psi_2 | \theta; \tau)
\right\} ].
\end{align*}
Here that $\varepsilon_1$ and $\varepsilon_2$ are independent $N(0,I)$ random vectors,
and $\psi_1, \psi_2$ are shorthand for
$g(\varepsilon_1, \theta, y)$ and $g(\varepsilon_2, \theta, y)$.

The final expression above allows Monte Carlo estimation of the required gradients of the objective $\mathcal{K}(\tau,A)$.

\paragraph{Optimisation algorithm}

We can now sketch an algorithm for optimal design, which is to iterate the following steps:
\begin{enumerate}
\item Sample $\theta, \psi, \epsilon$ given the current $\tau$.
\item Update $g$ by a step of SGD based on training data $(\theta, \psi, y(\epsilon, \theta, \tau))$.
\item Sample $\varepsilon_1, \varepsilon_2$.
\item Produce unbiased estimates of $\nabla_\tau \mathcal{K}$ and $\nabla_\eta \mathcal{K}$.
\item Update $\tau$ and $\eta$ by SGD steps.
\end{enumerate}
Upon convergence of $\tau$ to a limit point $\tau^*$,
$g$ will approximate the corresponding posterior for $\psi$.
Hence $\tau^*$ will approximate an optimal design.

\subsection{Intractable likelihood} \label{sec:ilikelihood}

Finally, consider the case where it is only possible to sample $y$ given $\theta, \tau$.
In this case we do not have access to an unbiased estimator of the score function.
However it is possible to train an estimate of the likelihood $\hat{f}(y | \theta)$ using neural density estimation methods.
This estimate can be used to define a score estimate $\hat{u}(y,\theta) = \nabla_\theta \log \hat{f}(y | \theta)$, allowing the approach of the previous subsection to be used.
The difference is that now $u$ is estimated purely from simulations,
while the previous section also made use of the full likelihood via \eqref{eq:Fisher_identity}.

\section{Poisson example derivations} \label{app:Poisson}

Here we derive the results stated in Section \ref{sec:illustration} on the Poisson example.
Recall that the design is $\tau \in [0,1]$,
and the observations are
$y_1 \sim Poisson(\tau \theta_1 \omega_1),
y_2 \sim Poisson((1-\tau) \theta_2 \omega_2)$,
and these are independent.
We assume $\omega_1 > \omega_2 > 0$ and that $\theta_1, \theta_2$ have independent $Gamma(2,1)$ priors.

\paragraph{Posterior distribution}

Using standard conjugacy relationships we have independent posterior distributions
$\theta_1 | y_1 \sim Gamma(2+y_1,1 + \tau \omega_1)$,
$\theta_2 | y_2 \sim Gamma(2+y_2,1 + (1-\tau) \omega_2)$.
Note that if $\tau=0$ then $y_1$ is a point mass at zero,
and so the $\theta_1$ posterior equals its prior.
Similarly if $\tau=1$ then the $\theta_2$ posterior equals its prior.

\paragraph{Expected Fisher information matrix}

For independent observations depending on different parameters, the FIM is diagonal with entries given by the scalar Fisher informations of the individual parameters (this is a straightforward consequence of the definition \eqref{eq:FisherInfo}).
Now consider $y \sim Poisson(\varphi)$.
The Fisher information with respect to $\varphi$ is $1/\varphi$.
Suppose $\varphi = k \vartheta$.
Using \eqref{eq:reparam}, the Fisher information with respect to $\vartheta$ is $k^2 / \varphi = k/\vartheta$.

Hence the FIM for this example is:
\[
\mathcal{I}_\theta(\theta) =
\begin{pmatrix}
\tau \omega_1 / \theta_1 & 0 \\ 0 & (1-\tau) \omega_2 / \theta_2
\end{pmatrix}.
\]
Note that $1/\theta_i$ has an inverse gamma (2,1) distribution which has expectation 1.
Thus
\[
\bar{\mathcal{I}}_\theta(\tau) =
\begin{pmatrix}
\tau \omega_1 & 0 \\ 0 & (1-\tau) \omega_2
\end{pmatrix}.
\]

\paragraph{ADV design}

We have $\det [\bar{\mathcal{I}}_\theta(\theta)] = \tau (1-\tau) \omega_1 \omega_2$ which is maximised by the design $\tau = 1/2$.
Furthermore we have already shown in Result \ref{res:GTreparam} that this design is invariant to linear reparameterisation.

\paragraph{FIG design}

We have $\trace [\bar{\mathcal{I}}_\theta(\theta)] = \tau \omega_1 + (1-\tau) \omega_2$.
Since $\omega_1 > \omega_2$, this is maximised by $\tau = 1$.

Next, consider a linear reparameterisation $\phi = B \theta$.
Then
\begin{align*}
\bar{\mathcal{I}}_\phi(\phi) &=
\E_{\theta \sim \pi(\theta)} [ \mathcal{I}_\phi(\phi) ] \\
&= \E_{\theta \sim \pi(\theta)} [B^{-T} \mathcal{I}_\theta(\theta) B^{-1}]
&& \text{from \eqref{eq:reparam}} \\
\Rightarrow
\trace \bar{\mathcal{I}}_\phi(\theta) &=
\trace [B^{-1} B^{-T} \bar{\mathcal{I}}_\theta(\theta) ]
&& \text{by linearity of expectation and} \\
&&& \text{\ cyclic property of trace} \\
& = \tau \omega'_1 + (1-\tau) \omega'_2
\end{align*}
where
$\begin{pmatrix} \omega'_1 \\ \omega'_2 \end{pmatrix}$
is the elementwise product of
$\diag(B^{-1} B^{-T})$ and $\begin{pmatrix} \omega_1 \\ \omega_2 \end{pmatrix}$.
The resulting optimal design is
\begin{itemize}
\item $\tau=1$ if $\omega'_1 > \omega'_2$
\item $\tau=0$ if $\omega'_1 < \omega'_2$
\item $\tau \in [0,1]$ if $\omega'_1 = \omega'_2$
\end{itemize}
and clearly $B$ can be chosen to allow all these possibilities.

\paragraph{SIG design}

Without loss of generality suppose $\omega_1=\omega_2=1$.
This is equivalent to a suitable reparameterisation of $\theta$, and the SIG design is invariant to reparameterisation.

For any $\tau < 1/2$, let $h = 1/2 - \tau$
and define
$X_1 \sim Poisson(\tau \theta)$,
$X_2 \sim Poisson(h \theta)$,
$X_3 \sim Poisson(h \theta)$.
Now let $S_i = \sum_{j=1}^i X_i$.
so that $S_1 \sim Poisson(\tau \theta)$,
$S_2 \sim Poisson(\theta/2)$,
$S_3 \sim Poisson([1-\tau] \theta)$.
Note that $X_1,X_2,X_3$ are not independent,
but are conditionally independent given $\theta$.

Define $v(\mathcal{A})$ as the expected Shannon information gain for parameter $\theta$ upon observing the set of variables $\mathcal{A}$.
Using sufficiency, we get $v(S_1) = v(X_1)$,
$v(S_2) = v(\{X_1,X_2\}) = v(\{X_1,X_3\})$
and $v(S_3) = v(\{X_1,X_2,X_3\})$.
We prove below that
\begin{equation} \label{eq:submod}
v(\{X_1,X_2\}) - v(X_1) > v(\{X_1,X_2,X_3\}) - v(\{X_1,X_3\}),
\end{equation}
so that $v(S_2) - v(S_1) > v(S_3) - v(S_2)$.
Rearranging gives $2v(S_2) > v(S_3) + v(S_1)$.

In our experiment we have two observations $Y_1, Y_2$, which are conditionally independent given $\theta$.
It follows from the definition of Shannon information gain that
$v(\{ Y_1,Y_2 \}) = v(Y_1) + v(Y_2)$.
Hence the expected Shannon information gain from the experiment
is $v(S_3) + v(S_1)$ under design $\tau$ (or $1-\tau$),
and $2 v(S_2)$ under design $1/2$.
Thus we have shown that the latter is optimal.

\paragraph{Proof of \eqref{eq:submod}}

Here we adapt the proof of proposition 2 from \cite{Krause:2005} (on submodularity of mutual information).
Define $\mathcal{X}$ to be a subset of $\{ X_1, X_2, X_3 \}$.
From Bayes theorem and conditional independence we have that
\begin{align*}
\log \pi(\theta | \mathcal{X}) - \log \pi(\theta)
&= \log \Pr(\mathcal{X}) + \log \Pr(\mathcal{X} | \theta) \\
&= \log \Pr(\mathcal{X}) + \sum_{X \in \mathcal{X}} \log \Pr(X | \theta).
\end{align*}
Taking the expectation with respect to $\theta$ and all $X$ variables gives
\[
v(\mathcal{X}) = H(\mathcal{X}) + \sum_{X \in \mathcal{X}} H(X'),
\]
where $H$ denotes Shannon entropy and the dash denotes conditioning on $\theta$.
Applying this result to both sides of \eqref{eq:submod} gives
\begin{align*}
v(\{X_1,X_2\}) - v(X_1) &= H(\{X_1,X_2\}) - H(X_1) + H(X'_2) \\
v(\{X_1,X_2,X_3\}) - v(\{X_1,X_3\}) &= H(\{X_1, X_2, X_3\}) - H(\{X_1,X_2\}) + H(X'_2).
\end{align*}
To prove \eqref{eq:submod} it remains to demonstrate the inequality
\[
H(\{X_1,X_2\}) - H(X_1) = H(X_2 | X_1) > H(X_2 | \{X_1, X_3\})
= H(\{X_1, X_2, X_3\}) - H(\{X_1,X_2\}).
\]
This holds due to the ``information can't hurt'' principle (see e.g.~\citealp{Cover:2012}, Theorem 2.6.5), with the non-independence of the $X_i$ variables giving strict inequality.

\section{Further details of pharmacokinetic example} 

\subsection{Fisher information matrix} \label{app:pkFIG}

The pharmacokinetic model is of the form $y \sim N(x(\theta,\tau), \sigma^2 I)$ with $\sigma$ fixed and $\dim x = n$.
Using \eqref{eq:mvn_info},
\begin{equation} 
\mathcal{I}_\theta(\theta) = \sigma^{-2} J^T J.
\end{equation}
where $J$ is a Jacobian matrix, with row $i$ column $j$ entry
$\frac{\partial x_i}{\partial \theta_j}$.
It remains to state these terms, which are
\begin{align*}
\frac{\partial}{\partial \theta_1} x_i(\theta, \tau)
&= \frac{1}{\theta_2 - \theta_1}x(\theta, \tau_i) - \frac{D\theta_2}{\theta_3(\theta_2 - \theta_1)}\tau_i \exp(-\theta_1\tau_i), \\
\frac{\partial}{\partial \theta_2} x_i(\theta, \tau)
&= \frac{\theta_1}{\theta_2(\theta_1-\theta_2)}x(\theta, \tau_i) + \frac{D\theta_2}{\theta_3(\theta_2 - \theta_1)}\tau_i \exp(-\theta_2\tau_i), \\
\frac{\partial}{\partial \theta_3} x_i(\theta, \tau)
&= - \frac{1}{\theta_3} x(\theta, \tau_i).
\end{align*}

\subsection{Choice of $K$} \label{app:Kcomparison}

\begin{figure}[htp]
\begin{center}
\includegraphics[width=0.45\textwidth]{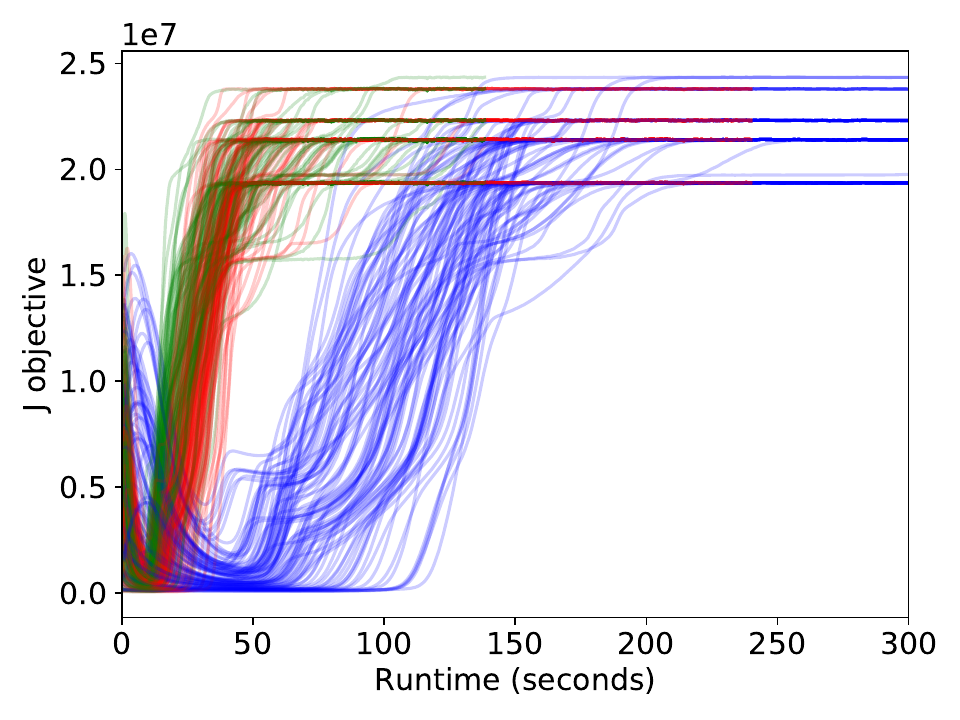}
\end{center}
\caption{
Trace plots of runtime versus $\mathcal{J}_{\text{ADV}}$ estimates for 100 replications of GDA on the pharmacokinetic example for $K=1$ (green), $K=10$ (red), $K=100$ (blue).
The runtime shown is the total runtime for all parallel replications.}
\label{fig:pk_Kcomparison}
\end{figure}

Section \ref{sec:pk_methods} mentions that,
when performing the GDA analysis of the pharmacokinetic example,
we compared several choices of $K$, the number samples to estimate $\mathcal{K}$ in \eqref{eq:Kest}.
Figure \ref{fig:pk_Kcomparison} shows the results of using $K=1,10,100$
by plotting $\mathcal{J}_{\text{ADV}}$ estimates, as defined in \eqref{eq:Jhat}.
It shows that convergence is slowest for $K=100$.
We use $K=1$ in the main paper as it appears slightly quicker than $K=10$.

\subsection{Results using \cite{Overstall:2017} approach} \label{app:pkACE}

Here we give some further details of our implementation of the pharmacokinetic example using the ACE method of \cite{Overstall:2017} for SIG optimisation.
The analysis presented in the main paper uses the default tuning choices in \cite{Overstall:2017}.
As mentioned in the main text, we experimented with one tuning choice but found little improvement.
The implementation uses the \texttt{acebayes} package \citep{Overstall:2017_R},
and our code can be found at
\url{https://github.com/dennisprangle/AdversarialDesignCode/R}.

The ACE software offers a second phase of point exchange optimisation.
Exploratory analysis showed that using this makes little change to the designs found for this example.
Therefore to avoid the extra computational cost, we did not use it for our main analysis.

\begin{figure}[htp]
\begin{center}
\includegraphics[width=0.45\textwidth]{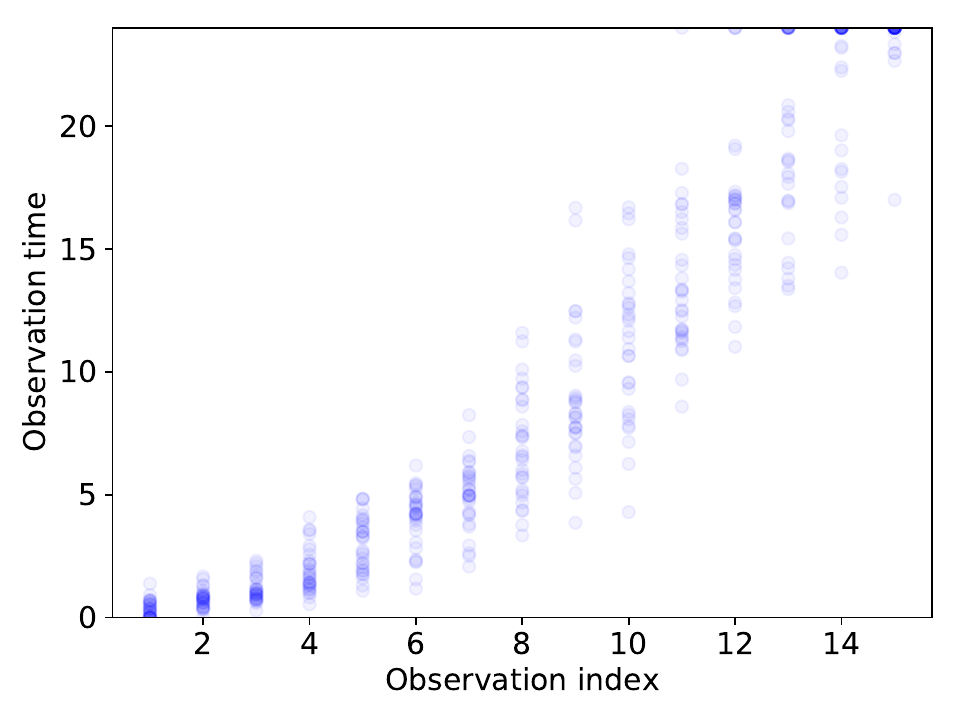}
\end{center}
\caption{Designs output by 30 runs of ACE for the pharmacokinetic example.
The horizontal axis shows the index of each point in the sorted design
i.e.~observation times are shown in increasing order from left to right.}
\label{fig:pk_ACE}
\end{figure}

Figure \ref{fig:pk_ACE} shows the designs produced for 30 runs from different initial designs.
The design points are spread out over the entire $[0,24]$ interval, although not uniformly.
This suggests that ACE has not completely converged,
as the SIG optimisation results using approaches from \cite{Foster:2020} (see next subsection)
are much less spread out.

\subsection{Results using \cite{Foster:2020} approaches} \label{app:pkPCE}

The main paper presents results for SIG optimisation for the pharmacokinetic example using methods from \cite{Foster:2020}.
An adaptation of their code to reproduce our analysis is available at \url{https://github.com/dennisprangle/pyro/tree/sgboed-reproduce}.
\cite{Foster:2020} propose several methods.
First we discuss prior contrastive estimation (PCE), as we found it converged most quickly for this application, and is therefore the method we report in the main paper to give a lower bound on the speed for methods from this paper.

\begin{figure}[htp]
\begin{center}
\includegraphics[width=0.45\textwidth]{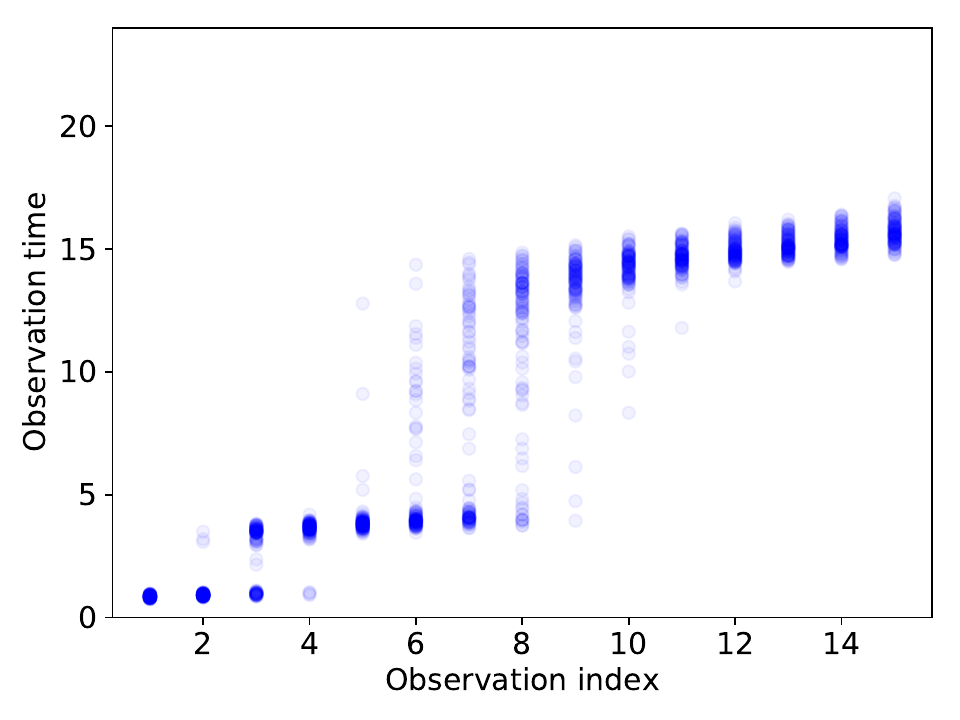}
\includegraphics[width=0.45\textwidth]{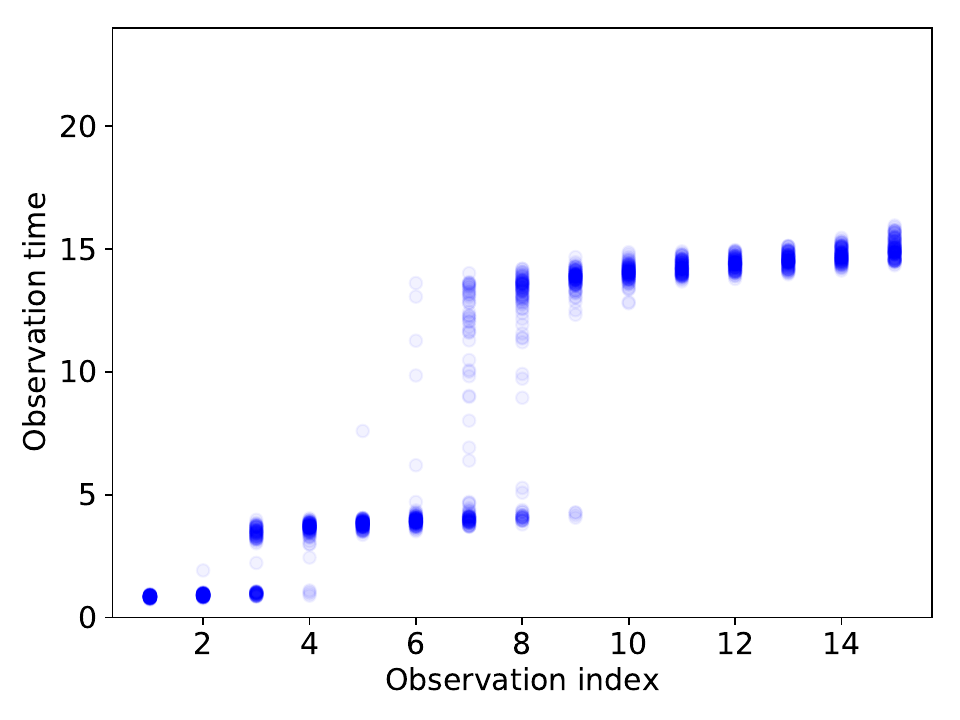}
\end{center}
\caption{Designs output by 100 runs of PCE for the pharmacokinetic example after 1 hour (left) and 2 hours (right).
The horizontal axis shows the index of each point in the sorted design
i.e.~observation times are shown in increasing order from left to right.}
\label{fig:pk_PCE}
\end{figure}

To investigate convergence of PCE we considered designs produced after 1 and 2 hours.
Figure \ref{fig:pk_PCE} shows these designs, for 100 replications of the optimisation algorithm, which were computed in parallel.
The designs after 2 hours are slightly different, suggesting that the method has not yet fully converged.
In the main paper we report results after 1 hour, taking this as a lower bound on the time required for convergence.

Figure \ref{fig:pk_PCE} has 3 clusters of design points, around\footnote{
We report these only to the nearest integer,
which is less precise than cluster locations for other methods elsewhere in the paper.
This is because the design points here are more variable across replications.}
$t=1,4,14$.
The points are more spread out around these clusters than the ADV results from the main paper, but this may be due to incomplete convergence, as discussed shortly.

\cite{Foster:2020} propose several alternatives to PCE,
and indeed PCE is not their overall recommendation.
For instance it is only guaranteed to converge to the optimal design in the limit of a large number of Monte Carlo samples being used to estimate gradients in each iteration.
However for this example we found the other methods were slower or less numerically stable, as we describe now.

We also attempted to implement the Barber-Agakov (BA) and adaptive contrastive estimation (ACE2\footnote{
We refer to this as ACE2 to distinguish it from the ACE method of \cite{Overstall:2017} discussed elsewhere in the paper.}) methods of \cite{Foster:2020}.
Both methods involve a neural network to map observations $y$ to an approximate posterior for $\theta$, $N(\mu(y), \Sigma(y))$.
We used a neural network with 2 fully connected layers of 16 hidden units and ReLU activation functions.
A final linear layer produces 9 outputs: 3 are used for $\mu$ and the other 6 for a Cholesky factor for $\Sigma(y)$.

\begin{figure}[htp]
\begin{center}
\includegraphics[width=0.45\textwidth]{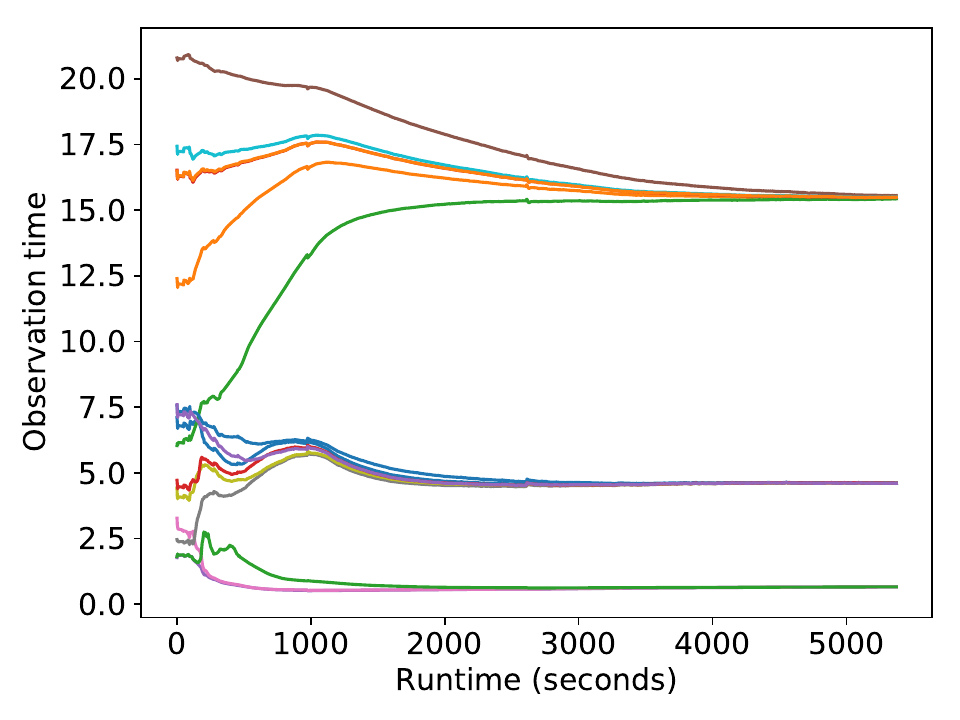}
\includegraphics[width=0.45\textwidth]{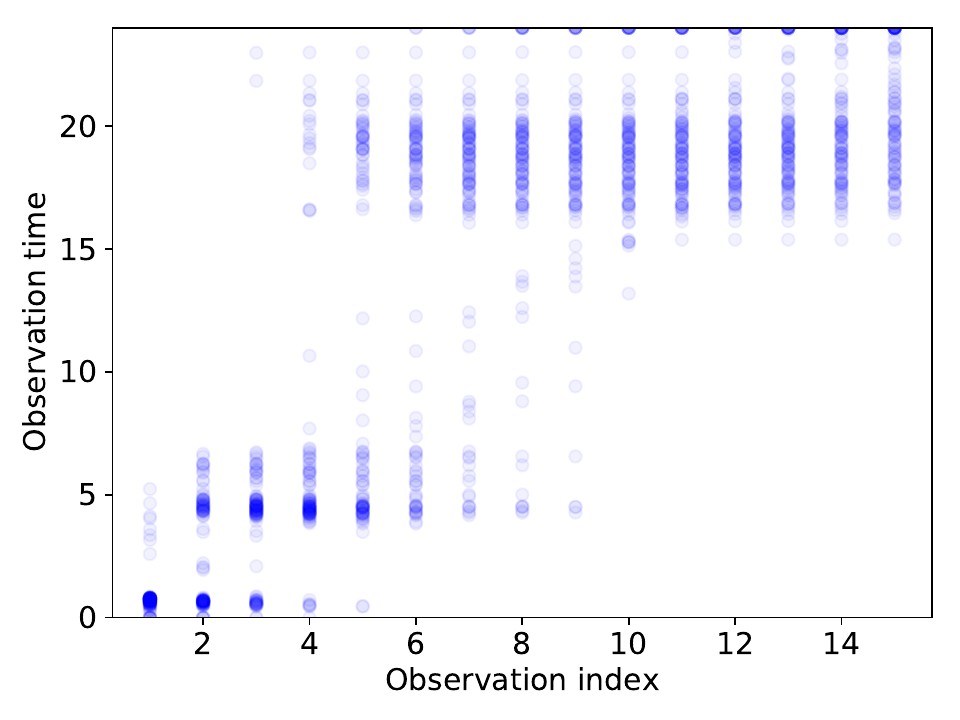}
\end{center}
\caption{Left: Trace plot for a single replication of BA on the pharmacokinetic example.
100 replications were run in parallel and the $x$-axis shows the total time to run all these replications.
Right: Designs output by all replications after 18 hours.
The horizontal axis shows the index of each point in the sorted design
i.e.~observation times are shown in increasing order from left to right.
}
\label{fig:pk_BA}
\end{figure}

We were unable to implement ACE2 as training produced NaN outputs.
So we discuss only the results of BA.
We ran 100 replications in parallel.
The left plot in Figure \ref{fig:pk_BA} shows the results of one well-behaved replication.
The design points converge to 3 clusters near $t=0.7,4.5,15.4$ after roughly 5,000 seconds (1.4 hours).
These are similar but not identical to the cluster locations for ADV reported in the main paper.
However many other replications did not converge so clearly to these locations, as shown by the right plot in Figure \ref{fig:pk_BA}.
Indeed, in a few replications we noticed the design remained very close to its initial values.

The BA runtime for the well-behaved replication is a little longer than the time used for PCE in the main paper -- which we ran for 1 hour (3,600 seconds) for 100 replications in parallel -- but of a similar order of magnitude.
However other replications did not appear to have converged after running for several hours.
As discussed by \cite{Foster:2020}, BA is likely to be somewhat slower than PCE as it needs to optimise a large number of neural network weights in addition to the design.

We expect that with further effort BA and ACE2 can be more successfully implemented for this example.
However, we believe that the PCE results reported in the main paper
capture a lower bound on the runtime for any of these methods on this example.

\subsection{SGD results} \label{app:pkSGD}

\begin{figure}[htp]
\begin{center}
\includegraphics[width=0.45\textwidth]{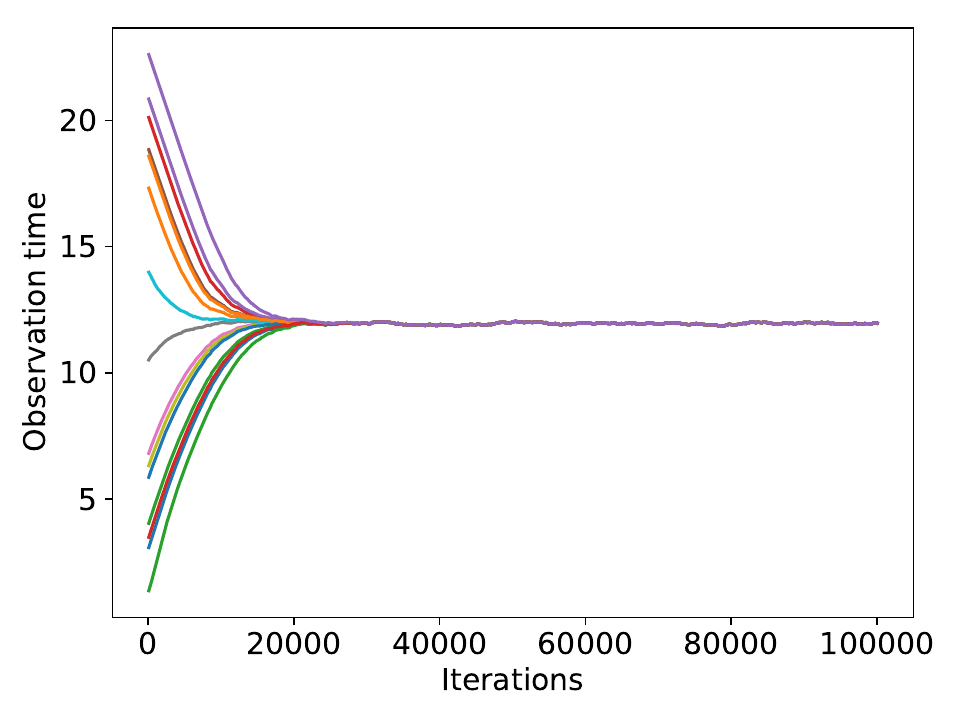}
\includegraphics[width=0.45\textwidth]{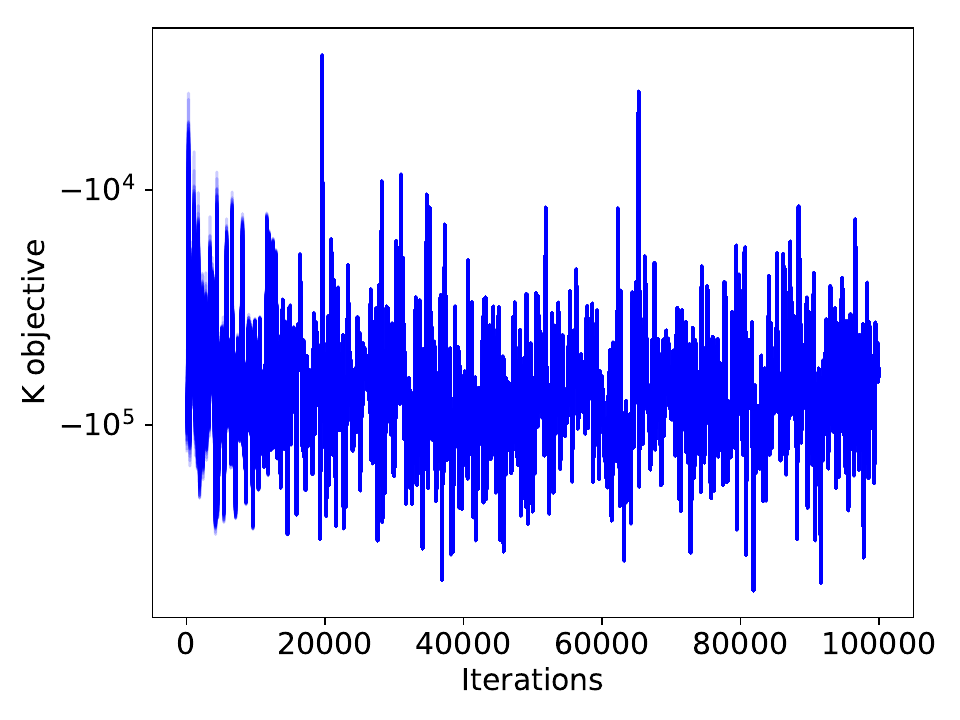} \\
\includegraphics[width=0.45\textwidth]{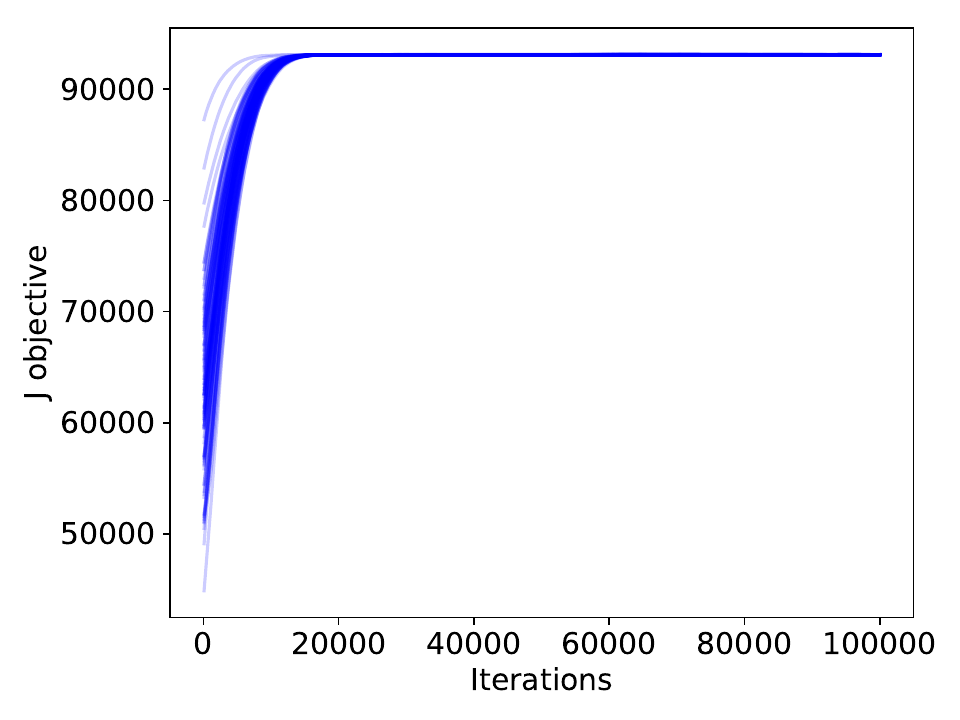}
\includegraphics[width=0.45\textwidth]{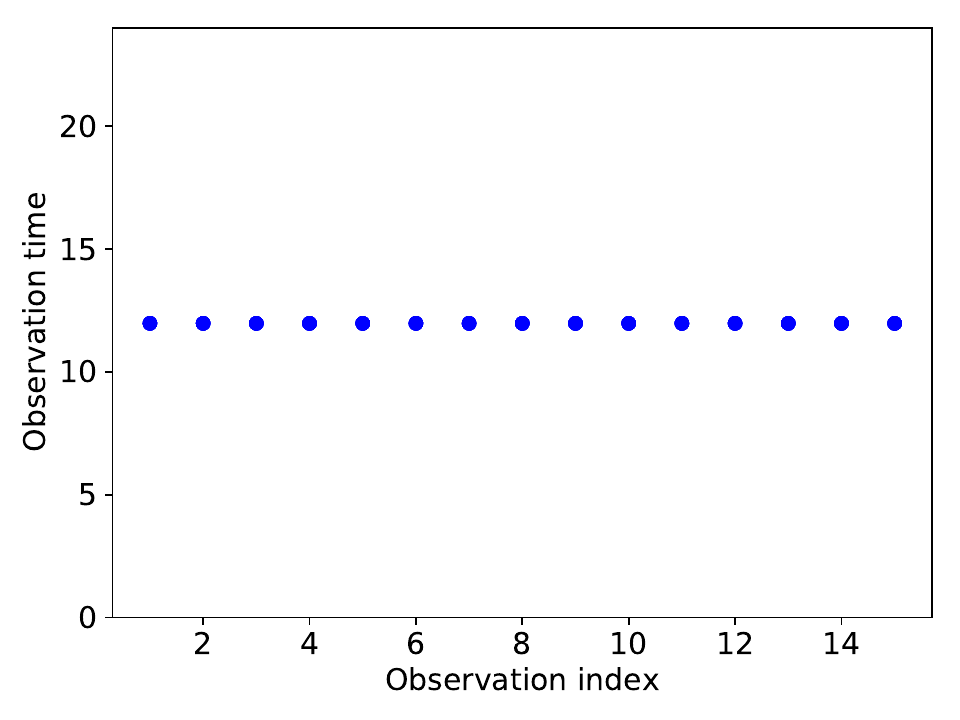}
\end{center}
\caption{Summary plots for FIG on the pharmacokinetic example.
Top left: designs during a single run of SGD.
Top right: $\mathcal{K}$ objective for 100 runs.
Bottom left: $\mathcal{J}_{\text{FIG}}$ objective for 100 runs.
Bottom right: designs output from 100 SGD runs.
The horizontal axis shows the index of each point in the sorted design
i.e.~observation times are shown in increasing order from left to right.
The designs are not identical between runs but the variation is too small to see on this scale.}
\label{fig:pk_SGD_traces}
\end{figure}

Here we consider using the FIG approach rather than ADV for the pharmacokinetic example.
To do so, Figure \ref{fig:pk_SGD_traces} shows results of optimising $\mathcal{J}_{\text{FIG}}$ using the SGD version of Algorithm \ref{alg:GDA_BED}.
The top left plot shows the evolution of the design during a single replication.
The design points converge to repeated observations at a single time.
The top right and bottom left plots display estimated $\mathcal{K}$ and $\mathcal{J}_{\text{FIG}}$ objectives over 100 repeated runs, which show rapid convergence.
The bottom right plot shows that in 100 runs, the design points always converge to a single cluster, around time $12$.
Point exchange was not implemented as multiple clusters were not observed.

\subsection{Shannon information gain} \label{app:pkSIG}

Figure \ref{fig:pk_boxplots} shows expected Shannon information gain values as a diagnostic for pharmacokinetic example design quality.
This subsection derives the calculations used for this diagnostic.

From \eqref{eq:expected utility} and \eqref{eq:SIG}, the expectation of the Shannon information gain utility is, in general
\[
\mathcal{J}_{\text{SIG}}(\tau) = \E_{(\theta,y) \sim \pi(\theta,y;\tau)} [ \log f(y | \theta; \tau) - \log \pi(y; \tau) ],
\]
where $f(y|\theta;\tau)$ is the likelihood and $\pi(y;\tau)$ is the evidence.

The pharmacokinetic model has the form $y = x(\theta, \tau) + \epsilon$ where $\epsilon \sim N(0,\sigma^2 \mathrm{I})$.
Let $\varphi(\epsilon)$ be the corresponding density.
Then:
\[
\E_{y \sim f(y|\theta;\tau)} [ \log f(y | \theta; \tau) ]
= \E_{\epsilon} [ \log \varphi(\epsilon) ]
= -H(\epsilon),
\]
where $H$ represents Shannon entropy.
Using the standard result for the Shannon entropy of the multivariate normal (see e.g.~\citealp{Cover:2012}, Theorem 8.4.1) gives
\[
H(\epsilon)
= \frac{\dim y}{2}[1 + \log(2\pi) + 2 \log \sigma]
= -13.25 \quad \text{(to 2 decimal places)}
\]
since $\dim y = 15$ and $\sigma = 0.1$.
Hence
\[
\mathcal{J}_{\text{SIG}}(\tau) = H(\epsilon) - \E_{(\theta,y) \sim \pi(\theta,y;\tau)} \log \pi(y; \tau).
\]
The second term can be estimated using a nested Monte Carlo approximation, to give
\begin{align*}
\hat{\mathcal{J}}_{\text{SIG}}(\tau)
&= H(\epsilon) - \frac{1}{B} \sum_{i=1}^B \log \hat{\pi}(y^{(i)}; \tau), \\
\text{where} \quad \hat{\pi}(y; \tau)
&= \frac{1}{B'} \sum_{j=1}^{B'} f(y | \theta^{(j)}; \tau).
\end{align*}
Here the $y^{(i)}$s are independent samples from the prior predictive under design $\tau$, and the $\theta^{(j)}$s are independent samples from the prior.
To reduce cost and variance, we reuse the same $\theta$ and $\epsilon$ samples
in all evaluations of $\hat{\mathcal{J}}_{\text{SIG}}(\tau)$.
Following \cite{Overstall:2017} we take $B=B'=1000$.

\section{Realistic pharmacokinetic example} \label{sec:realistic_supp}

Here we describe the analysis used for the realistic version of the pharmacokinetic example described in Section \ref{sec:pk_realistic} of the main paper.
Section \ref{sec:realistic_methods} describes the changes in methodology required,
and Section \ref{sec:realistic_discussion} gives further discussion of the results, which were presented in the main paper.

\subsection{Methodology} \label{sec:realistic_methods}

\paragraph{Gaps}

We wish to impose the constraint that observations must be separated by at least $0.25$ hours.
To do so it was sufficient to follow the approach described in Section \ref{sec:main alg} and simply add the following penalty term to our objective:
\[
\sum_{i=1}^{14} 1000 \max(0, 0.25 - [\tau_{(i+1)} - \tau_{(i)}])
\]
where $\tau_{(i)}$ is the $i$th observation time after sorting into increasing order.
Thus a penalty is added whenever there are gaps between observation times less than $0.25$ hours.

\paragraph{Multiplicative noise}

We also wish to use the multiplicative noise model \eqref{eq:multi} i.e.
\[
y_i \sim N(x(\theta, \tau_i), \sigma_1^2 + \sigma_2^2 x(\theta, \tau_i)^2).
\]
This is a situation where the likelihood is available,
so we use Algorithm \ref{alg:GDA_BED2}
as described in Section \ref{sec:intractable_score}.

Let $\epsilon$ be a $N(0,I)$ random vector of length 30.
Then observations from the multiplicative noise model can be expressed as
\[
y_i(\epsilon,\theta,\tau) = x(\theta, \tau_i) + \sigma_1 \epsilon_i
+ \sigma_2 x(\theta, \tau_i) \epsilon_{i+15},
\]
as required by assumption A6,
and needed to implement Algorithm \ref{alg:GDA_BED2}.

We found analyses using multiplicative noise take longer to converge.
The likely explanation for this is that under the multiplicative noise the gradient estimates are more variable,
since the Fisher information matrix must now be estimated.
Therefore we use $500,000$ iterations here,
rather than the $100,000$ iterations used for the other pharmacokinetic analyses.
These analyses took approximately half an hour to run,
roughly 10 times as long as the analyses without multiplicative noise.
Also, as discussed in Section \ref{sec:intractable_score}, in this time
the multiplicative noise analyses perform only a single replication of experimental design, while the others perform 100 replications in parallel.

\subsection{Discussion} \label{sec:realistic_discussion}

Figure \ref{fig:pk_designs} in the main paper shows designs found from four analyses,
under all combinations of with/without gaps and with/without multiplicative noise.
The design with multiplicative noise and gaps is similar to those found in previous work using SIG: see Figure 2c in \citealp{Overstall:2017}.

Close inspection of Figure \ref{fig:pk_designs} shows that the design points for multiplicative noise without gaps are slightly more spaced out than those without multiplicative noise.
Inspection of trace plots suggest this is due to lack of full convergence of the optimisation algorithm.
As mentioned above, this is likely due to more variable gradient estimates.

Each row of Figure \ref{fig:pk_designs} shows a design from a single run of a GDA optimisation algorithm.
Therefore each gives a local optimum of our ADV objective (or a close approximation to one).
As in the other pharmacokinetic examples,
we expect that there are multiple local optima which differ in the number of design points in each cluster.
Selecting between these would require further work,
such as applying point exchange when no gaps are used,
or developing a similar post-processing method for the case with gaps.

\section{Further details of geostatistical regression example}
\label{app:geostats}

\begin{figure}[htp]
\begin{center}
\includegraphics[width=0.45\textwidth]{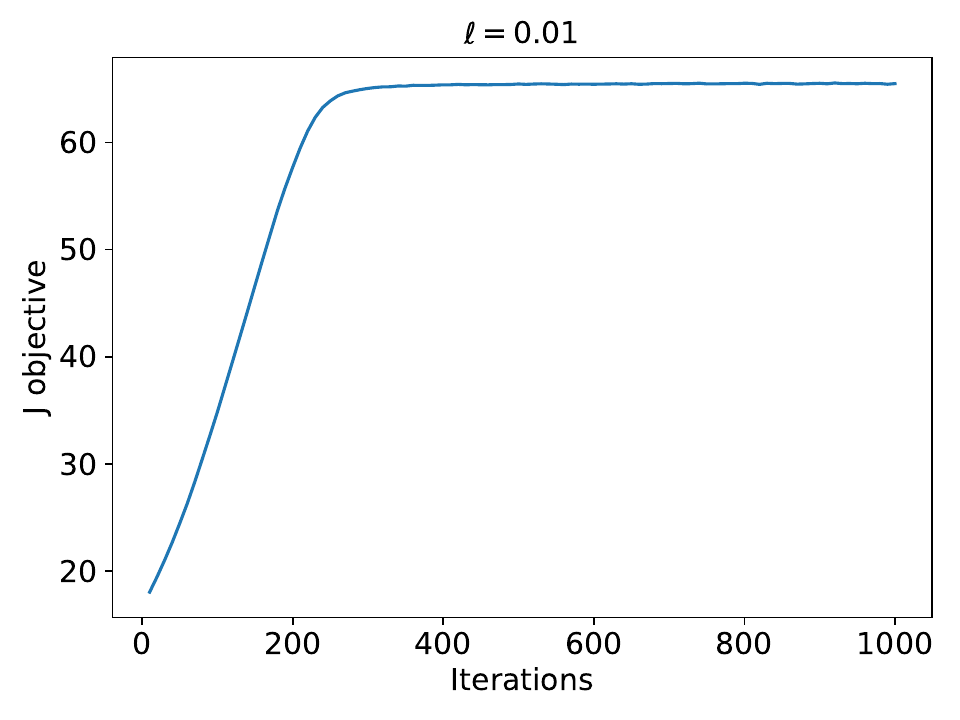}
\includegraphics[width=0.45\textwidth]{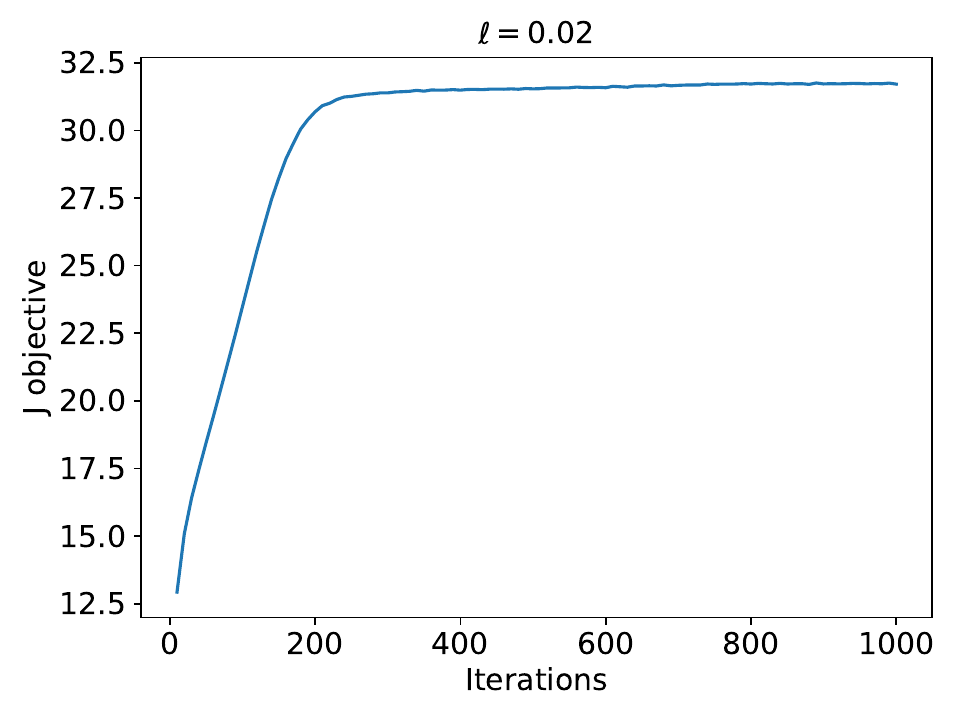} \\
\includegraphics[width=0.45\textwidth]{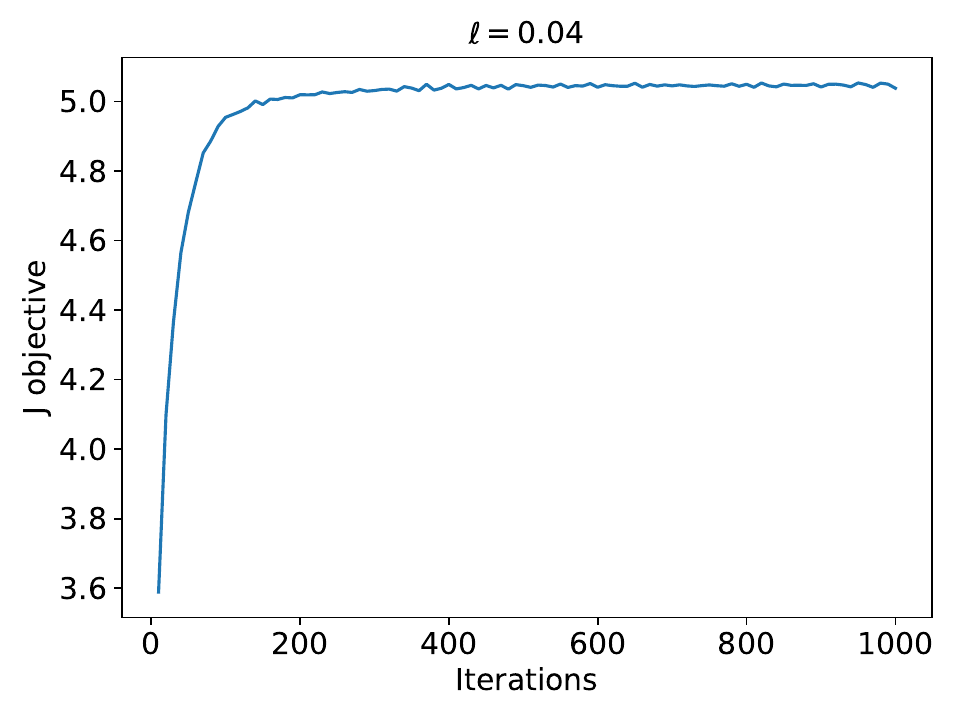}
\includegraphics[width=0.45\textwidth]{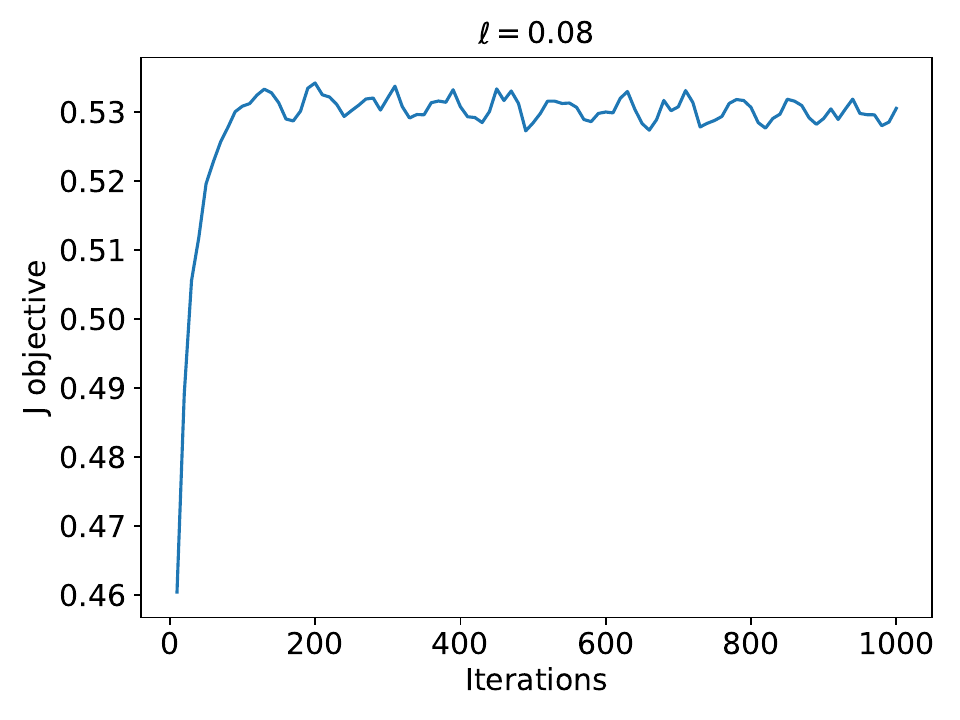}
\end{center}
\caption{Trace plots of $\mathcal{J}_{\text{ADV}}$ estimates for the geostatistical regression example.}
\label{fig:geostats_J_traces}
\end{figure}

Figure \ref{fig:geostats_J_traces} shows trace plots of $\mathcal{J}_{\text{ADV}}$ estimates,
confirming convergence after 1000 iterations.
It also shows that the improvement in this objective is largest for small $\ell$.

\end{document}